\documentclass[]{JHEP3}
\usepackage{epsfig}

\def\simlt{\stackrel{<}{{}_\sim}}

\def\be{\begin{equation}}
\def\ee{\end{equation}}
\def\beq{\begin{equation}}
\def\eeq{\end{equation}}
\def\bea{\begin{eqnarray}}
\def\eea{\end{eqnarray}}

\title{Neutrino Mass\footnote{Review article
submitted for publication in Contemporary Physics}}

\author{S. F. King \footnote{e-mail address: sfk@hep.phys.soton.ac.uk}\\
Department of Physics and Astronomy,
University of Southampton, Southampton, SO17 1BJ, U.K.}


\abstract{This is a review article about the most recent
developments on the field of neutrino mass.
The first part of the review introduces the idea of neutrino masses
and mixing angles, summarizes the most recent experimental
data then discusses the experimental prospects and challenges in this
area. The second part of the review discusses the implications
of these results for particle physics and cosmology,
including the origin of neutrino mass, the see-saw mechanism and
sequential dominance, and large extra dimensions and cosmology.
}

\preprint{SHEP 07-21}

\begin{document}

\section{Introduction}

\footnote{There are plenty of good reviews of neutrino physics, and here are just a few of them:
\cite{Mohapatra:2006gs,Mohapatra:2005wg,King:2003jb,Altarelli:2004za,Mohapatra:2003qw,bilenky,Murayama:2002xw,Bahcall:2004cc}.
This Introduction is intended to be a rapid overview of the subject,
and much of the material contained here will be explained in greater
depth in the body of this review.}
In 1930, the Austrian physicist Wolfgang Pauli proposed the existence
of particles called neutrinos, denoted as $\nu$,
as a ``desperate remedy'' to account for
the missing energy in a type of radioactivity called beta decay. At
the time physicists were puzzled because nuclear beta decay appeared
to violate energy conservation. In beta decay, a neutron in an
unstable nucleus transforms into a proton and emits an electron, where
the radiated electron was found to have a continuous energy
spectrum. This came as a great surprise to many physicists because
other types of radioactivity involved gamma rays and alpha particles
with discrete energies.  Pauli deduced that some of the energy must
have been taken away by a new particle emitted in the decay process,
the neutrino, which carries energy and has spin 1/2, but which is
massless, electically neutral and very weakly interacting.
Because neutrinos interact so
weakly with matter, Pauli bet a case of champagne that nobody would
ever detect one, and they became known as ``ghost particles''.
Indeed it was not until a quarter of a century later, in
1956, that Pauli lost his bet and neutrinos were discovered when Clyde
Cowan and Fred Reines detected antineutrinos emitted from a nuclear
reactor at Savannah River in South Carolina, USA.

\begin{figure}[t]
\centering
\includegraphics[width=0.75\textwidth]{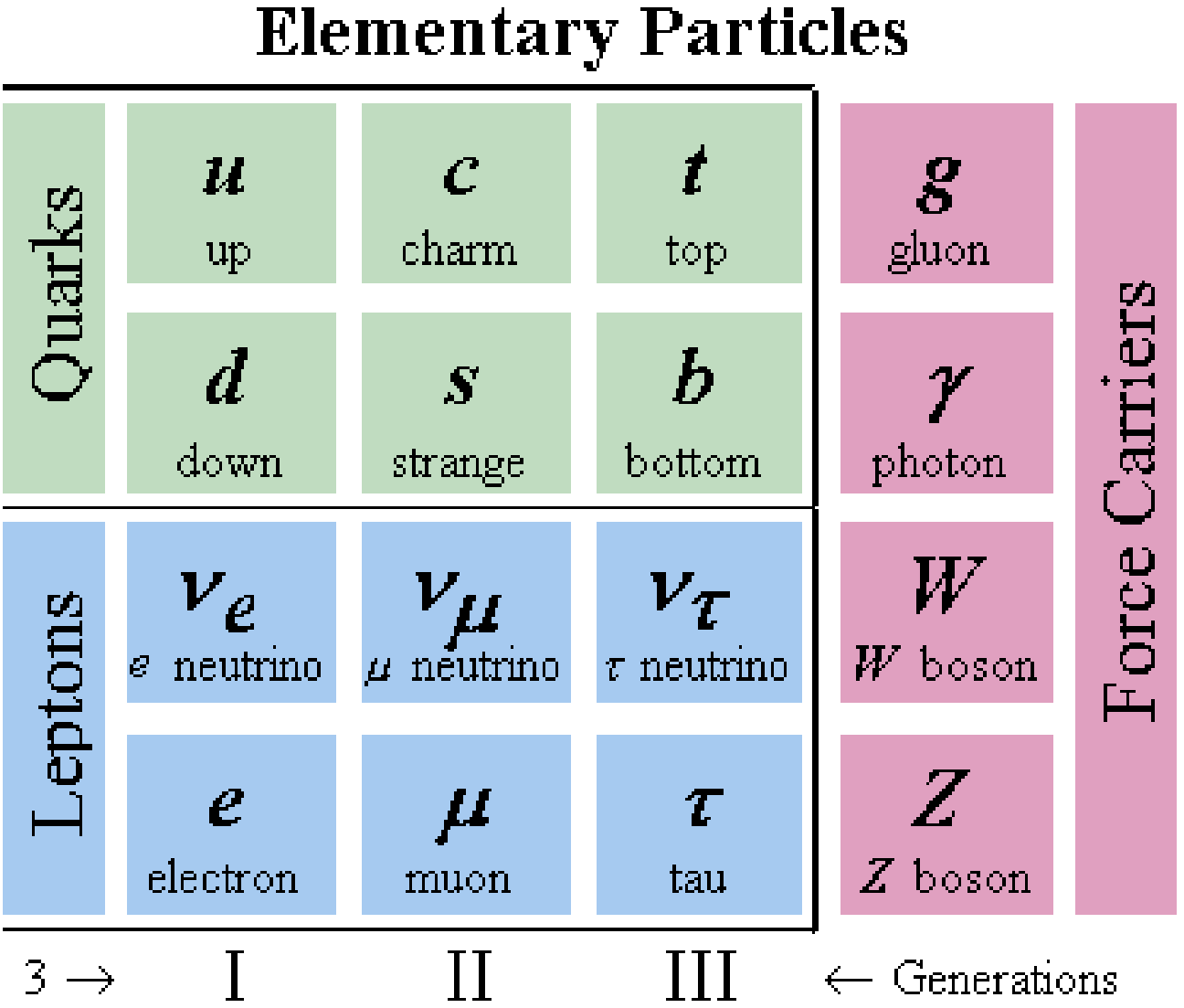}
\vspace*{-4mm}
    \caption{The particles of the Standard Model.}
\label{constituents}
\vspace*{-2mm}
\end{figure}

Since then, after decades of painstaking experimental and theoretical work,
neutrinos have become enshrined as an essential part of the accepted
quantum description of fundamental particles and forces, the Standard
Model of Particle Physics,
whose particle content is summarized in Fig.\ref{constituents}.
This is a highly successful theory in which
elementary building blocks of matter are divided into three generations
of two kinds of particle - quarks and leptons. It also includes three
of the fundamental forces of Nature, the strong ($g$), electromagnetic
($\gamma$) and weak ($W,Z$) forces carried by spin 1 force carrying
bosons (shown in parentheses) but does not include gravity.
There are six flavours of quarks given in Fig.\ref{constituents}.
The leptons consist of three flavours of
charged leptons, the electron $e^-$, muon $\mu^-$
and tau $\tau^-$, together with
three flavours of neutrinos - the electron neutrino $\nu_e$, muon
neutrino $\nu_{\mu}$ and tau neutrino $\nu_{\tau}$ which are our main
concern here.

The first clues that neutrinos have mass
came from an experiment deep underground, carried out
by an American scientist Raymond Davis Jr., detecting solar
neutrinos \cite{Bahcall:2004cc}.
It revealed only about one-third of the number predicted by theories of
how the Sun works pioneered by John Bahcall \cite{Bahcall:2004cc}.
The result puzzled both solar and neutrino
physicists. However, some Russian researchers, Mikheyev and Smirnov,
developing ideas proposed previously by Wolfenstein in the U.S.,
suggested that the solar neutrinos might be changing into something
else.  Only electron neutrinos are emitted by the Sun and they could
be converting into muon and tau neutrinos which were not being
detected by the experiments.
This effect called ``neutrino oscillations'', as the
types of neutrino interconvert over time from one kind to another, was
first proposed some time earlier by Pontecorvo
\cite{Pontecorvo:1957qd}. The precise mechanism
for ``solar neutrino oscillations''
proposed by Mikheyev, Smirnov and Wolfenstein involved the resonant
enhancement of neutrino oscillations due to matter effects.
Just as light passing through matter slows down,
which is equivalent to the photon gaining a small effective mass,
so neutrinos passing through matter also result in the neutrinos
slowing down and gaining a small effective mass.
The effective neutrino mass is largest when the matter density
is highest, which in the case of solar neutrinos is in the core
of the Sun. In particular electron neutrinos generated in the core of the Sun will
be subject to such matter effects. It turns out that
neutrino oscillations, which would be present in the vacuum
due to neutrino mass and mixing,
will exhibit strong resonant effects in the presence of matter
as the effective mass of the neutrinos varies along the path length
of the neutrinos. This can result in a resonant enhancement of solar neutrino
oscillations known as the MSW effect \cite{MSW}.

\begin{figure}[t]
\includegraphics[width=0.76\textwidth]{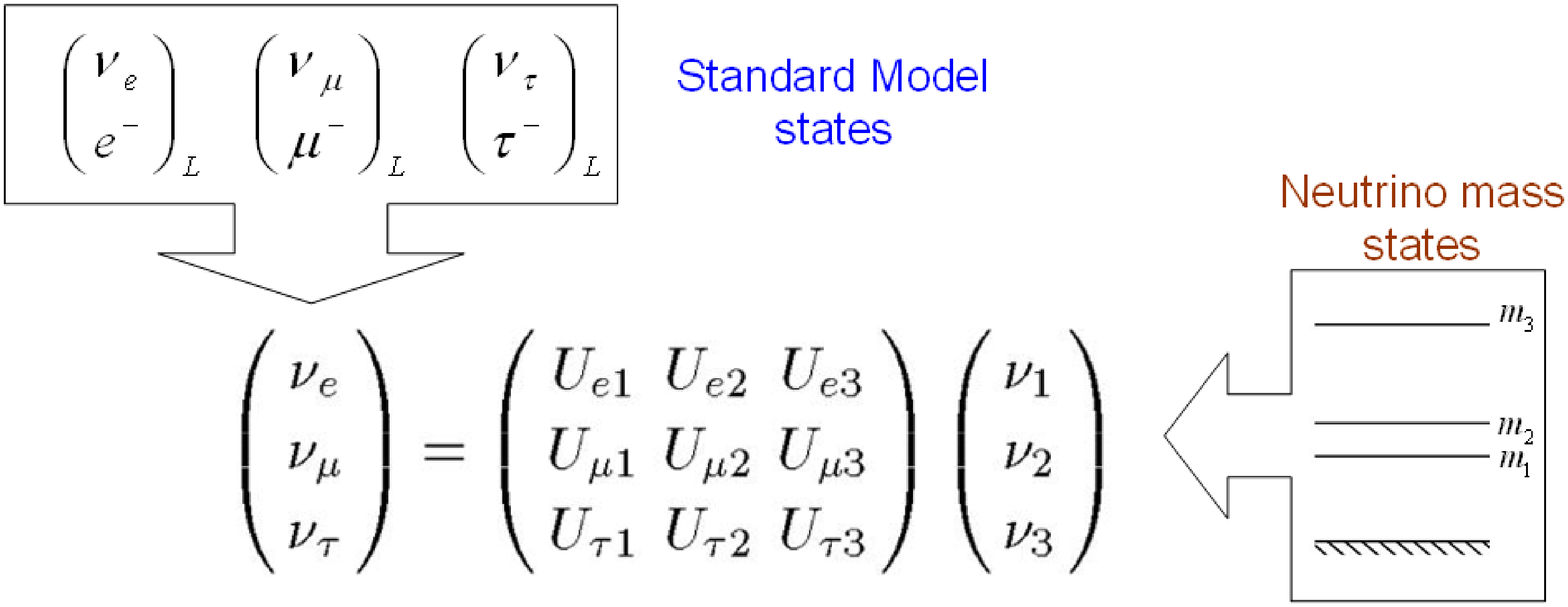}
\vspace*{-4mm}
    \caption{The lepton mixing matrix
$U$ relates Standard Model neutrino states
$\nu_e$, $\nu_{\mu}$, $\nu_{\tau}$ to the
neutrino mass states $\nu_1$, $\nu_2$, and $\nu_3$
with mass $m_1$, $m_2$, and $m_3$.
The mass states are not necessarily ordered as
$m_1<m_2<m_3$, as discussed later. The convention
is chosen such that $\nu_1$ contains mostly $\nu_e$,
while $\nu_3$ contains mainly $\nu_{\mu}$
and $\nu_{\tau}$, with very little $\nu_e$.
The state $\nu_2$ contains roughly equal amounts of $\nu_e$,
$\nu_{\mu}$ and $\nu_{\tau}$.
The matrix $U$ is unitary, which implies that the probability that
each of the states $\nu_1$, $\nu_2$, and $\nu_3$ contains each of
$\nu_e$, $\nu_{\mu}$, $\nu_{\tau}$ must sum to unity.
} \label{MNS1}
\vspace*{-2mm}
\end{figure}

Neutrino oscillations are analagous to coupled pendulums, where
oscillations in one pendulum induce oscillations in another pendulum.
The coupling strength is defined in terms of something called the
``lepton mixing matrix'' $U$
\footnote{The ``lepton mixing matrix'' $U$ is also frequently
referred to as the Maki-Nakagawa-Sakata (MNS) matrix $U_{MNS}$
\cite{MNS}, and sometimes the name of Pontecorvo is added at the beginning
to give $U_{PMNS}$.}
which relates the basic Standard Model
neutrino states, $\nu_e$, $\nu_{\mu}$, $\nu_{\tau}$,
associated with the electron, muon and tau,
to the neutrino mass states $\nu_1$, $\nu_2$, and $\nu_3$
with mass $m_1$, $m_2$, and $m_3$, as shown in Fig.\ref{MNS1}.
According to quantum mechanics it is not necessary that the Standard
Model states $\nu_e$, $\nu_{\mu}$, $\nu_{\tau}$ be identified in a one-one
way with the mass eigenstates $\nu_1$, $\nu_2$, and $\nu_3$,
and the matrix elements of $U$ give the quantum amplitude that
a particular Standard Model state contains an admixture of a
particular mass eigenstate. As with all quantum amplitudes,
the matrix elements of $U$ are expected to be complex numbers in
general.

The idea of
neutrino oscillations gained support from the Japanese experiment
Super-Kamiokande \cite{SK}
which in 1998 showed that there was a deficit of muon
neutrinos reaching Earth when cosmic rays strike the upper atmosphere,
the so called ``atmospheric neutrinos''. Since most neutrinos pass
through the Earth unhindered, Super-Kamiokande was able to detect
muon neutrinos coming from above and below, and found that while the
correct number of muon neutrinos came from above, only about a half
of the expected number came from below. The
results were interpreted as half the
muon neutrinos from below oscillating into tau
neutrinos over an oscillation length $L$ of the diameter of the Earth,
with the muon neutrinos from above having a negligible oscillation
length, and so not having time to oscillate, yielding the
expected number of muon neutrinos from above.
More recently, the Sudbury Neutrino Observatory (SNO) in Canada
has spectacularly confirmed the ``solar neutrino oscillations''
\cite{Ahmed:2003kj}. The
experiment measured both the flux of the electron neutrinos and the
total flux of all three types of neutrinos. The SNO data revealed that
physicists' theories of the Sun were correct after all, and the solar
neutrinos $\nu_e$ were produced at the standard rate but were
oscillating into $\nu_{\mu}$ and $\nu_{\tau}$, with only about a third
of the original $\nu_e$ flux arriving at the Earth.
Since then, neutrino oscillations consistent with solar
neutrino observations
have been seen using man made neutrinos from nuclear
reactors at KamLAND in Japan \cite{Eguchi:2002dm,Maricic:2005sg},
and neutrino oscillations consistent with atmospheric
neutrino observations have been seen using neutrino
beams fired over hundreds of kilometers
as in the K2K experiment in Japan
\cite{Ahn:2006zz},
the Fermilab-MINOS experiment in the US \cite{Michael:2006rx}
or the CERN-OPERA experiment in Europe.
Further long-baseline neutrino beam experiments are in the pipeline,
and neutrino oscillation physics is poised to enter the precision era,
with Superbeams and a Neutrino Factory on the horizon.

Following these results
several research groups \cite{GoPe,Maltoni:2004ei}
showed that the electron
neutrino has a mixing matrix element of $|U_{e2}|\approx 1/\sqrt{3}$
which is the quantum amplitude for $\nu_e$ to contain
an admixture of the mass eigenstate $\nu_2$ corresponding to
a massive neutrino of mass
$m_2\approx 0.007$ electronvolts (eV) or greater
(by comparison the electron has a mass of about half a
megaelectronvolt (MeV)).
The muon and tau neutrinos were observed to contain
approximately equal amplitudes of a heavier neutrino
$\nu_3$ of mass $m_3\approx 0.05$ eV or greater,
$|U_{\mu 3}|\approx |U_{\tau 3}| \approx 1/\sqrt{2}$,
where a normalized amplitude of $1/\sqrt{2}$ corresponds
to a 1/2 fraction of $\nu_3$ in each of $\nu_{\mu}$ and $\nu_{\tau}$,
leading to a maximal mixing and oscillation of
$\nu_{\mu} \leftrightarrow \nu_{\tau}$.
However, according to the results from the CHOOZ nuclear reactor
experiment \cite{Apollonio:1999ae},
the electron neutrino must only mix very weakly
(if at all) with this state, $|U_{e3}|<0.2$.
Neutrino oscillations are only sensitive to mass differences,
and the lightest neutrino mass $m_1$ is not measured,
so these mass values are only lower bounds.
However, as discussed later,
there are cosmological reasons to believe that none of the neutrino
masses can exceed about 0.3 eV.
Clearly, then,  neutrino masses are much smaller than the other charged
fermion masses, and this represents something of a puzzle. However
there is a more urgent question that must be faced since,
unlike the case for quarks and charged leptons, the Standard
Model actually predicts that neutrinos have no mass at all!

The most intuitive way to understand why neutrino mass is forbidden
in the Standard Model, is to understand that
the Standard Model predicts that neutrinos
always have a ``left-handed'' spin - rather like rifle bullets which
spin counter clockwise to the direction of travel.
In fact this property was
first experimentally measured in 1958,
two years after the neutrino was discovered, by
Maurice Goldhaber, Lee Grodzins and Andrew Sunyar.  More accurately,
the ``handedness'' of a particle describes the direction of its spin vector
along the direction of motion, and the neutrino being ``left-handed''
means that its spin vector always points in the opposite direction to its
momentum vector.  The fact that the neutrino is left-handed, written as
$\nu_L$, implies that it
must be massless. If the neutrino has mass then, according to special
relativity, it can never travel at the speed of light. In principle,
a fast moving observer could therefore overtake the
spinning massive neutrino and would see it moving in the opposite
direction. To the observer, the massive neutrino would therefore
appear right-handed. Since the Standard Model predicts that neutrinos must
be strictly left-handed, it follows that neutrinos are massless
in the Standard Model. It also follows that the discovery of neutrino
mass implies new physics Beyond the Standard Model, with
profound implications for particle physics and cosmology.

The rest of the review is organized as follows.
In section \ref{mass} neutrino masses and mixing angles
will be defined more precisely, assuming no prior knowledge,
and starting with two neutrino mixing, building up to three
neutrino mixing, eventually with complex CP violating phases.
Also the the current experimental
status and future prospects will be discussed in some more detail.
In section \ref{implications} the implications of neutrino mass
for particle physics and cosmology are described,
including the origin of neutrino mass, the see-saw mechanism and
sequential dominance, and large extra dimenensions and cosmology.
Finally section \ref{conclusions} concludes the review.

\section{Neutrino Masses and Mixing Angles}
\label{mass}

The history of neutrino oscillations dates back to the work of
Pontecorvo who in 1957 \cite{Pontecorvo:1957qd}
proposed $\nu \rightarrow \bar{\nu}$ oscillations in
analogy with $K \rightarrow \bar{K}$ oscillations,
described as the mixing of two Majorana neutrinos.
Majorana neutrinos will be explained later in this review,
but for now it is sufficient to define them as neutrinos
which are equivalent to their own antiparticles.
Pontecorvo was the first to
realise that what we call the ``electron neutrino'', for example,
may be a linear combination of mass eigenstate neutrinos, and that
this feature could lead to neutrino oscillations
of the kind $\nu_e \rightarrow \nu_{\mu}$. Later on MSW proposed that such
neutrino oscillations could be resonantly enhanced in the Sun
\cite{MSW}. The present section introduces the basic formalism
of neutrino masses and mixing angles, gives an up-to-date
summary of the current experimental status of this fast moving field,
and discusses future experimental prospects.

\subsection{Two state atmospheric neutrino mixing}
\label{twostate}
In 1998 the Super-Kamiokande experiment published a
paper \cite{SK}
which represents a watershed in the history of neutrino physics.
The Super-Kamiokande experiment consists of
thousands of tonnes of pure water in a tank deep underground,
and was originally built to search for proton decay.
However, its designers realized that the
experiment might also be able to detect highly energetic neutrinos
from the Sun that interact with electrons via scattering reactions.
These electrons can travel faster than the local speed of light in the water,
causing them to emit the optical equivalent of a sonic boom
 - a glow of blue light called Cerenkov radiation that
can be detected by ultra-sensitive photomultiplier tubes around the tank.
Super-Kamiokande also measured the number of
electron and muon neutrinos that arrive at the Earth's surface as a
result of cosmic ray interactions in the upper atmosphere, which are
referred to as ``atmospheric neutrinos''. While the number and and
angular distribution of electron neutrinos is as expected,
Super-Kamiokande showed that the number of muon neutrinos is
significantly smaller than expected and that the flux of muon
neutrinos exhibits a strong dependence on the zenith angle. These
observations gave compelling evidence that muon neutrinos undergo
flavour oscillations and this in turn implies that at least one
neutrino flavour has a non-zero mass. The standard interpretation,
well supported by current data, is
that muon neutrinos are oscillating into tau neutrinos.

\begin{figure}
\begin{center}
\includegraphics[width=12.0cm]{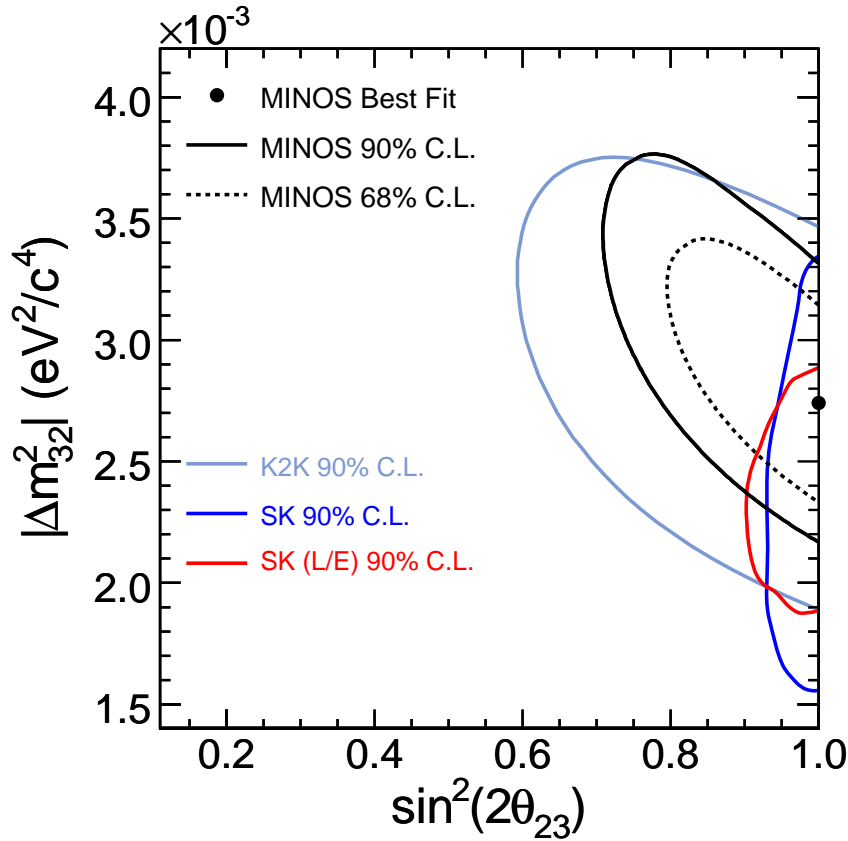}
\caption{\label{fig:MINOS-Oscillation-Contours-With-K2K-Super-K}
Confidence intervals from the MINOS experiment \cite{Michael:2006rx}.
Results from K2K and Super-Kamiokande (SK) are also shown.}
\end{center}
\end{figure}

Current atmospheric neutrino oscillation data are well described
by simple two-state mixing
\begin{equation}
\left(\begin{array}{c} \nu_\mu \\ \nu_\tau \end{array} \\ \right)=
\left(\begin{array}{cc}
 \cos \theta_{23} & \sin \theta_{23} \\
 -\sin \theta_{23} & \cos \theta_{23} \\
\end{array}\right)
\left(\begin{array}{c}
\nu_2 \\ \nu_3 \end{array} \\ \right)
\; ,
\end{equation}
and the two-state probability oscillation formula
\be
P(\nu_{\mu }\rightarrow \nu_{\tau})=\sin^22\theta_{23}
\sin^2(1.27\Delta m_{32}^2{L}/{E})
\ee
where
\begin{equation}
\Delta m^2_{ij} \equiv m^2_i - m^2_j \;
\end{equation}
and $m_i$ are the physical neutrino mass eigenvalues associated
with the mass eigenstates $\nu_i$.
$\Delta m_{32}^2$ is in units of eV$^2$, the baseline $L$ is in km and
the beam energy $E$ is in GeV.
The atmospheric data
results support maximal mixing, with best-fit two-neutrino oscillation
parameters of
\beq
\sin^2 2\theta_{23} = 1, \ \
\Delta m_{32}^2 = 2.6 \times 10^{-3}
\mathrm{eV}^2.
\eeq
The 90\% C.L. range for $\Delta m_{32}^2$ at $\sin^2 2\theta_{23} = 1$ is
between 2.0 and 3.2 $\times 10^{-3} \mathrm{eV}^2$.
The experimental results for such neutrino oscillations are
usually plotted as confidence level contours
in the $\Delta m_{32}^2$-$\sin^2 2\theta_{23}$
plane as shown in
Fig.\ref{fig:MINOS-Oscillation-Contours-With-K2K-Super-K}.
The results are dominated by the latest SuperKamiokande results,
but the recent results from the long baseline neutrino beam experiments
K2K \cite{Ahn:2006zz} and MINOS \cite{Michael:2006rx}
are also shown on the same plot.

The approximately maximal mixing angle $\theta_{23}=45^\circ$ means that
we identify the heavy atmospheric neutrino of mass $m_3$ as being approximately
\beq
\nu_3\approx \frac{\nu_{\mu}+\nu_{\tau}}{\sqrt{2}}
\eeq
and in addition there is a lighter orthogonal combination of mass $m_2$,
where $m_3^2-m_2^2=2.6 \times 10^{-3}\ eV^2$.
If $m_3\gg m_2$ then this implies $m_3\approx 0.05 \ eV$.

\subsection{Three family neutrino mixing}

Super-Kamiokande is also sensitive to the electron neutrinos arriving
from the Sun, the ``solar neutrinos'', and has independently confirmed
the reported deficit of such solar neutrinos long reported by other
experiments.  For example Davis's Homestake Chlorine experiment
which began data taking in 1970 consists of 615 tons of
tetrachloroethylene, and uses radiochemical techniques to determine
the Ar$^{37}$ production rate \cite{Bahcall:2004cc}.
More recently the SAGE and Gallex
experiments contain large amounts of Ga$^{71}$ which is converted to Ge71
by low energy electron neutrinos arising from the dominant pp reaction
in the Sun \cite{Bahcall:2004cc}.
The combined data from these and other experiments
implies an energy dependent suppression of solar neutrinos which can
be interpreted as due to flavour oscillations. Taken together with the
atmospheric data, this requires that a second neutrino flavour has a
non-zero mass.

SNO is a water Cerenkov detector like Super-Kamiokande,
but instead of using normal water it uses heavy water, $D_2O$.
The deuterons, D, in the heavy water are the most weakly bound of all nuclei,
which gives SNO the chance
to observe three different reactions induced by solar neutrinos.
The first of these processes is the charged-current (CC)
reaction $\nu_e + D \rightarrow  p + p + e^-$,
which is detected by observing Cerenkov photons from
the energetic recoil electron, $e^-$.
SNO also measures
the neutral-current (NC) reaction $\nu_{\alpha} + D \rightarrow p + n + \nu_{\alpha}$.
This is observed via the emitted neutrons, n,
and is independent of the flavour of the incoming neutrino, $\nu_{\alpha}$.
It therefore provides
a way to normalize the total flux of neutrinos being emitted by the Sun.
Finally SNO measures the elastic scattering (ES) reaction also measured in Super-Kamiokande,
$\nu_{\alpha} + e^- \rightarrow \nu_{\alpha} + e^-$,
which has some sensitivity to all neutrino flavours.
SNO measurements of CC reaction on deuterium is
sensitive exclusively to $\nu_e$'s, while the ES
off electrons also has a small sensitivity to $\nu_{\mu}$'s and
$\nu_{\tau}$'s.  The CC ratio is significantly smaller than the ES
ratio. This immediately disfavours oscillations of $\nu_e's$ to
sterile neutrinos \footnote{Sterile neutrinos are defined to
be a light neutrino with no weak interactions.}
which would lead to a diminished flux of electron
neutrinos, but equal CC and ES ratios.  On the other hand the
different ratios are consistent with oscillations of $\nu_e$'s to
active neutrinos $\nu_{\mu}$'s and $\nu_{\tau}$'s since this would
lead to a larger ES rate since this has a neutral current
component. The SNO analysis is nicely consistent with both the
hypothesis that electron neutrinos from the Sun oscillate into other
active flavours, and with the Standard Solar Model prediction. The
latest results from SNO including the data taken with salt inserted
into the detector to boost the efficiency of detecting the neutral
current events \cite{Ahmed:2003kj},
strongly favour the large solar mixing angle (LMA) MSW
solution, discussed more below.  In other words there is no longer any solar neutrino
problem: we have instead solar neutrino mass!

The minimal neutrino sector required to account for the
atmospheric and solar neutrino oscillation data thus consists of
three light physical neutrinos with left-handed flavour eigenstates,
$\nu_e$, $\nu_\mu$, and $\nu_\tau$, defined to be those states
that share the same doublet as the charged lepton mass eigenstates
$e,\mu ,\tau$
(see Fig.\ref{MNS1}).
Within the framework of three--neutrino oscillations,
the neutrino flavor eigenstates $\nu_e$, $\nu_\mu$, and $\nu_\tau$ are
related to the neutrino mass eigenstates $\nu_1$, $\nu_2$, and $\nu_3$
with mass $m_1$, $m_2$, and $m_3$, respectively, by a $3\times3$
unitary matrix called the lepton mixing matrix $U$
\cite{MNS}
\begin{equation}
\left(\begin{array}{c} \nu_e \\ \nu_\mu \\ \nu_\tau \end{array} \\ \right)=
\left(\begin{array}{ccc}
U_{e1} & U_{e2} & U_{e3} \\
U_{\mu1} & U_{\mu2} & U_{\mu3} \\
U_{\tau1} & U_{\tau2} & U_{\tau3} \\
\end{array}\right)
\left(\begin{array}{c} \nu_1 \\ \nu_2 \\ \nu_3 \end{array} \\ \right)
\; .
\label{MNS0}
\end{equation}

If the light neutrinos are Majorana,
$U$ can be parameterized in terms of three mixing angles
$\theta_{ij}$ and three complex phases.
A unitary matrix has six phases but three of them are removed
by the phase symmetry of the charged lepton Dirac masses.
Since the neutrino masses are Majorana there is no additional
phase symmetry associated with them, unlike the case of quark
mixing where a further two phases may be removed.

\begin{figure}[t]
\centering
\includegraphics[width=0.76\textwidth]{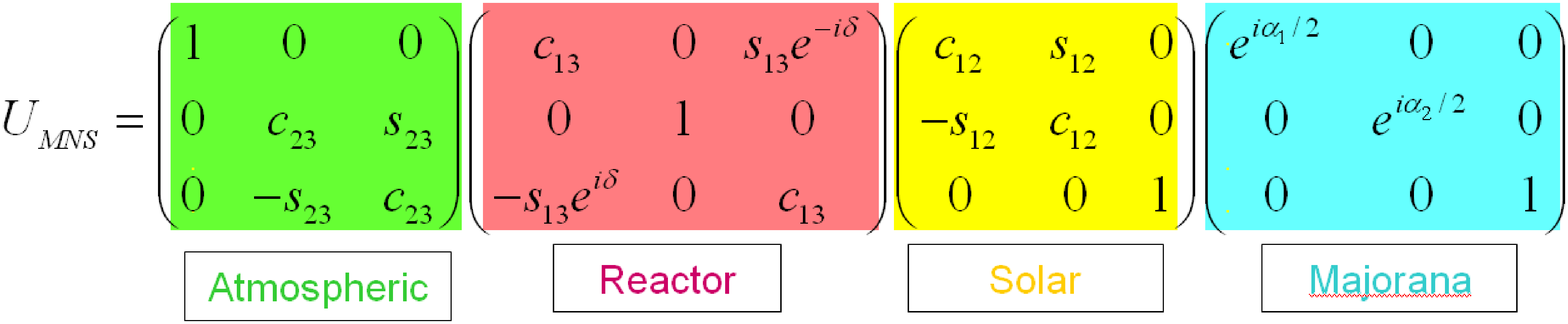}
\vspace*{-4mm}
    \caption{The lepton mixing matrix with phases factorizes
into a matrix product of four matrices,
associated with the physics of Atmospheric neutrino oscillations,
Reactor neutrino oscillations, Solar neutrino oscillations
and a Majorana phase matrix.} \label{MNS2}
\vspace*{-2mm}
\end{figure}

If we begin by assuming that the phases are zero, then
the lepton mixing matrix may be parametrised by
a product of three Euler rotations, as depicted in
Fig.\ref{fig2}, and given by a product of three matrices:
\begin{equation}
U=R_{23}R_{13}R_{12}
\label{euler}
\end{equation}
where
\begin{equation}
R_{23}=
\left(\begin{array}{ccc}
1 & 0 & 0 \\
0 & c_{23} & s_{23} \\
0 & -s_{23} & c_{23} \\
\end{array}\right), \ \
R_{13}=
\left(\begin{array}{ccc}
c_{13} & 0 & s_{13} \\
0 & 1 & 0 \\
-s_{13} & 0 & c_{13} \\
\end{array}\right), \ \
R_{12}=
\left(\begin{array}{ccc}
c_{12} & s_{12} & 0 \\
-s_{12} & c_{12} & 0\\
0 & 0 & 1 \\
\end{array}\right)
\end{equation}
where $c_{ij} = \cos\theta_{ij}$ and $s_{ij} = \sin\theta_{ij}$.
Note that the allowed range of the angles is
$0\leq \theta_{ij} \leq \pi/2$.
Including phases, the lepton mixing matrix is summarized in
Fig.\ref{MNS2}. The phases $\alpha_{1,2}$ are called Majorana
phases since they are only present if the neutrino mass is
Majorana (as defined earlier and discussed later). The phase $\delta$ is called the Dirac phase
since it is always present even if neutrinos have Dirac mass.
We have already seen that the first matrix in Fig.\ref{MNS2}
is associated with Atmospheric neutrino oscillations.
We now discuss the physics associated with the other
matrix factors.

%
\FIGURE[h]{
\label{fig2}
 \unitlength=1in
\begin{picture}(4.5,2.7)
\put(-0.2,0){\epsfig{file=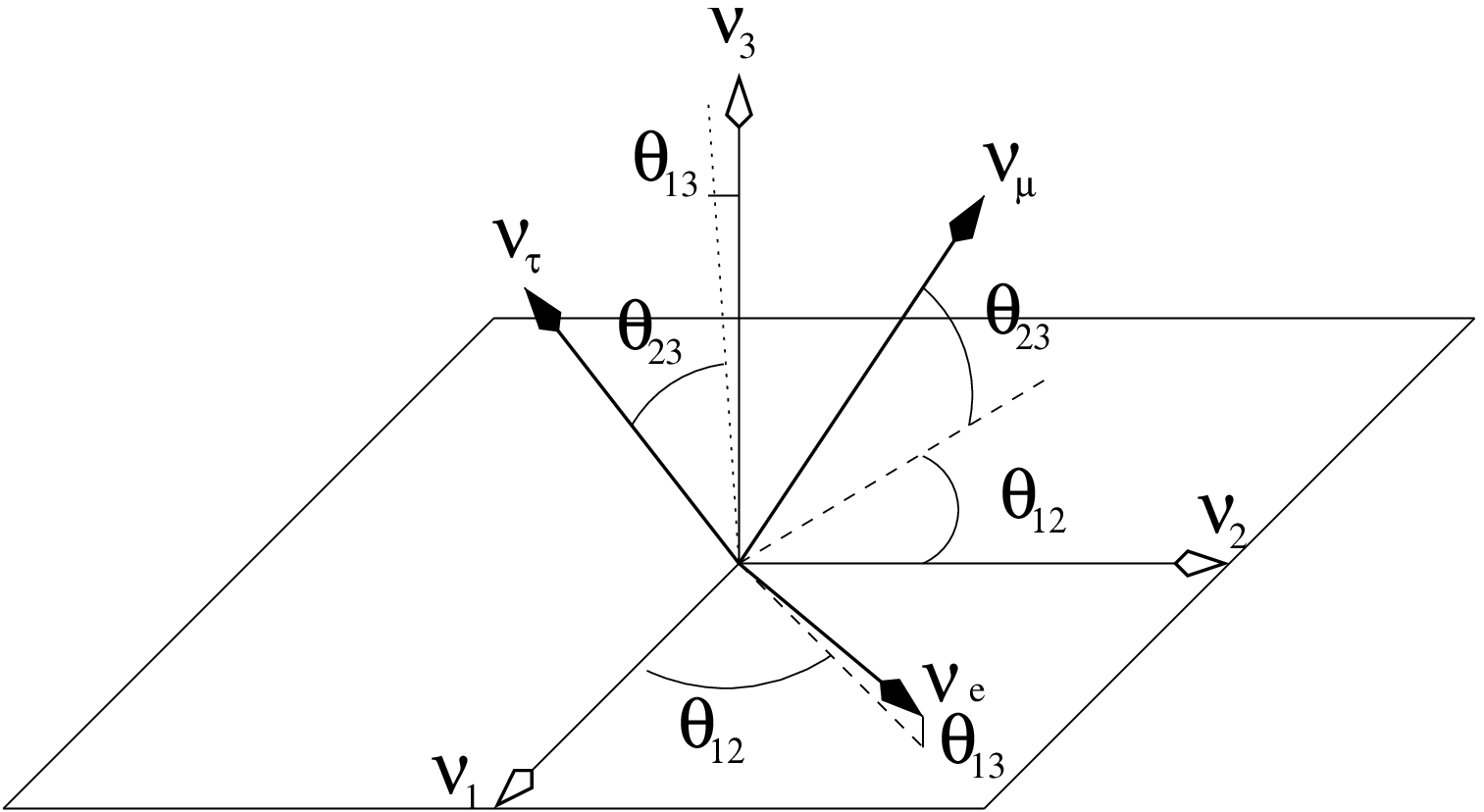, width=5in}}
\end{picture}
\caption
{The relation between
the neutrino weak eigenstates $\nu_e$, $\nu_\mu$, and $\nu_\tau$ and
the neutrino mass eigenstates $\nu_1$, $\nu_2$, and $\nu_3$
in terms of the three mixing angles $\theta_{12}$,
$\theta_{13}$, $\theta_{23}$.
Ignoring phases, these are just the Euler angles
respresenting the rotation of one orthogonal basis
into another.}
}

The physics of the second matrix in Fig.\ref{MNS2} is associated
with Reactor neutrino oscillations.
Reactor experiments detect the anti-electron neutrinos
which are produced copiously in the cores of nuclear reactors,
and interpret any deficit in the expected number
of such particles in terms of neutrino oscillations.
The solar neutrino background is low because the Sun
produces electron neutrinos, with negligible numbers of anti-electron neutrinos.
The CHOOZ reactor experiment in France failed to see any signal of
anti-neutrino oscillations over the Super-Kamiokande mass range.  CHOOZ
data from $\bar{\nu}_{e}\rightarrow \bar{\nu}_{e}$ disappearance not
being observed provides a significant constraint on $\theta_{13}$ over
the Super-Kamiokande (SK) prefered range of $\Delta m_{32}^2$
\cite{Apollonio:1999ae}:
\begin{equation}
\sin^2 \theta_{13}<0.04
\end{equation}
The CHOOZ experiment therefore limits $\sin \theta_{13} \simlt 0.2$
or $\theta_{13} \simlt 12^\circ$
over the favoured atmospheric range at 90\% C.L.
The experiment is currently being upgraded to Double
CHOOZ, to increase the sensitivity on the angle $\theta_{13}$.
The phase $\delta$ also appears in the third matrix,
and physically represents CP violation
(see section 3.1 for a discussion of CP violation).
Since the angle
$\theta_{13}$ that it is associated with has not yet been
measured it might seem
somewhat premature to discuss the phases associated with this
angle. Nevertheless there is in fact a huge experimental effort under way
to both measure the angle $\theta_{13}$ and the CP phase
$\delta$. However it should be emphasised that  the $CP$-violation
in the lepton sector is one of the most
challenging frontiers in the future studies
of neutrino mixing. Nevertheless the experimental searches
for $CP$-violation in neutrino oscillations
can help answer the fundamental question about
the status of CP-symmetry in the
lepton sector at low energy. The observation of
leptonic $CP$-violation at low energies
will have far reaching consequences, and can shed light, in particular,
on the possible origin of the baryon asymmetry
of Universe.

The physics of the third matrix in Fig.\ref{MNS2} is associated
with Solar neutrino oscillations, as discussed above,
and recently confirmed by the Japanese reactor experiment
KamLAND, that measures
$\bar{\nu}_e$'s produced by several surrounding nuclear
reactors \cite{Maricic:2005sg}.
KamLAND has already seen
a signal of neutrino oscillations over the Solar neutrino
LMA MSW mass range,
and has recently confirmed the LMA MSW region ``in the laboratory''
\cite{Eguchi:2002dm}.
KamLAND and SNO results when combined
with other solar neutrino data especially that of Super-Kamiokande
uniquely specify the large mixing angle (LMA) MSW \cite{MSW} solar solution
with three active light neutrino states, a large solar angle
\beq
\sin^2 \theta_{12} \approx 0.30, \ \
\Delta m_{21}^2\approx 7.9\times 10^{-5}{\rm eV}^2.
\eeq
according to the most recent global fits \cite{Maltoni:2004ei}.
KamLAND has thus not only confirmed solar neutrino oscillations,
but has also uniquely specified the large mixing angle (LMA) solar
solution, heralding a new era of precision neutrino physics.

The physics of the fourth matrix in Fig.\ref{MNS2} is associated
with Majorana neutrino masses. These phases could in principle
be measured in neutrinoless double beta decay, discussed later.

\begin{figure}[t]
\centering
\includegraphics[width=0.56\textwidth]{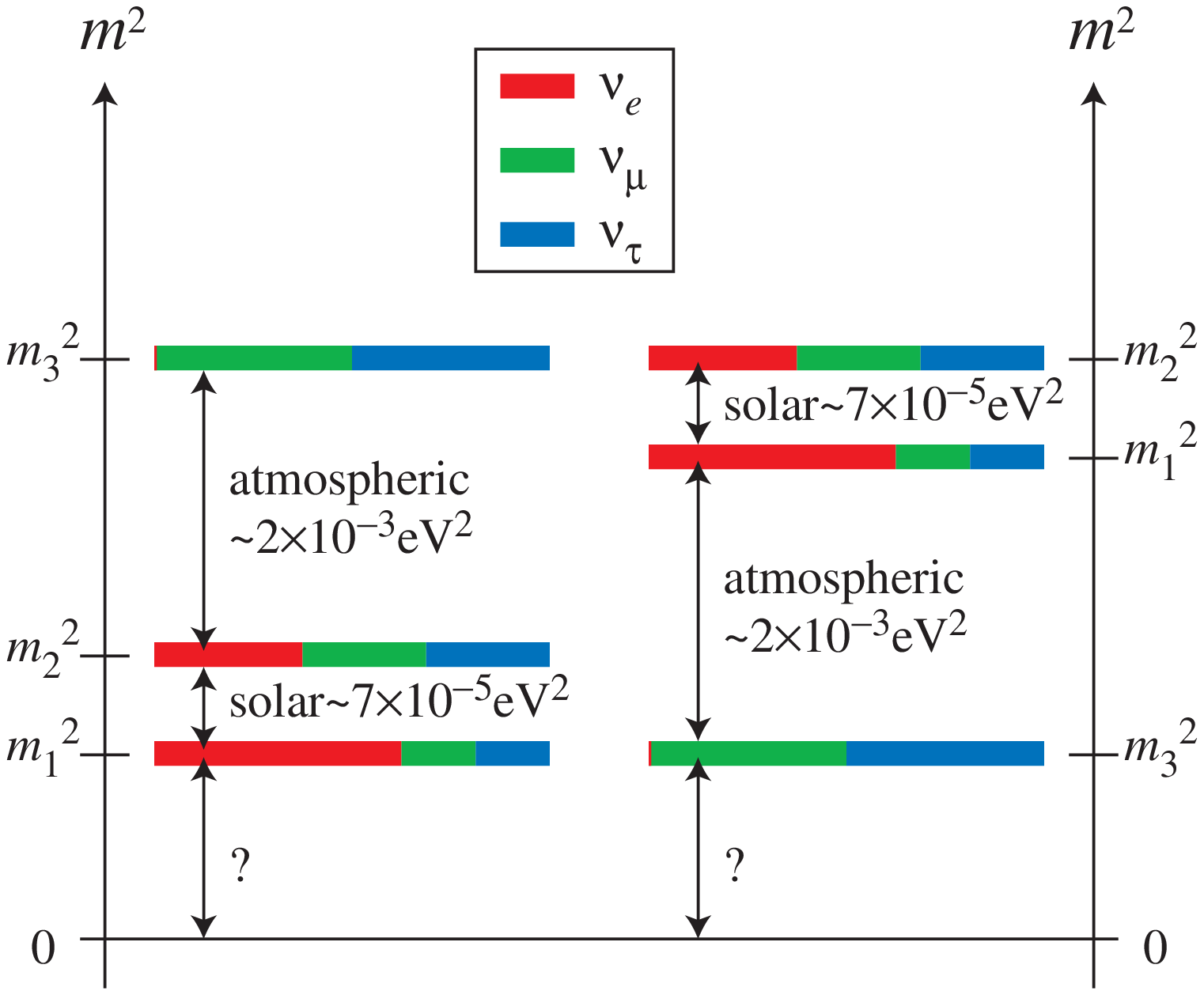}
\vspace*{-4mm}
    \caption{Alternative neutrino mass
patterns that are consistent with neutrino
oscillation explanations of the atmospheric and solar data.
The pattern on the left (right) is called the normal (inverted)
pattern. The coloured bands represent the probability
of finding a particular weak eigenstate
$\nu_e$, $\nu_\mu$, and $\nu_\tau$
in a particular mass eigenstate.
The absolute scale of neutrino masses is not
fixed by oscillation data
and the lightest neutrino mass may vary from $0.0-0.3$ eV.}
\label{fig1}
\vspace*{-2mm}
\end{figure}

It is clear that neutrino oscillations, which
only depend on $\Delta m_{ij}^2\equiv m_i^2-m_j^2$,
give no information about the absolute value of the neutrino mass squared
eigenvalues $m_i^2$,
and there are basically two
patterns of neutrino mass squared orderings
consistent with the atmospheric and solar data as shown in
Fig.\ref{fig1}. Three family
oscillation probabilities depend upon the time--of--flight (and hence the
baseline $L$), the $\Delta m^2_{ij}$, and $U$ (and hence
$\theta_{12}, \theta_{23}, \theta_{13}$, and $\delta$).

\begin{figure}
\centering
\includegraphics[width=0.56\textwidth]{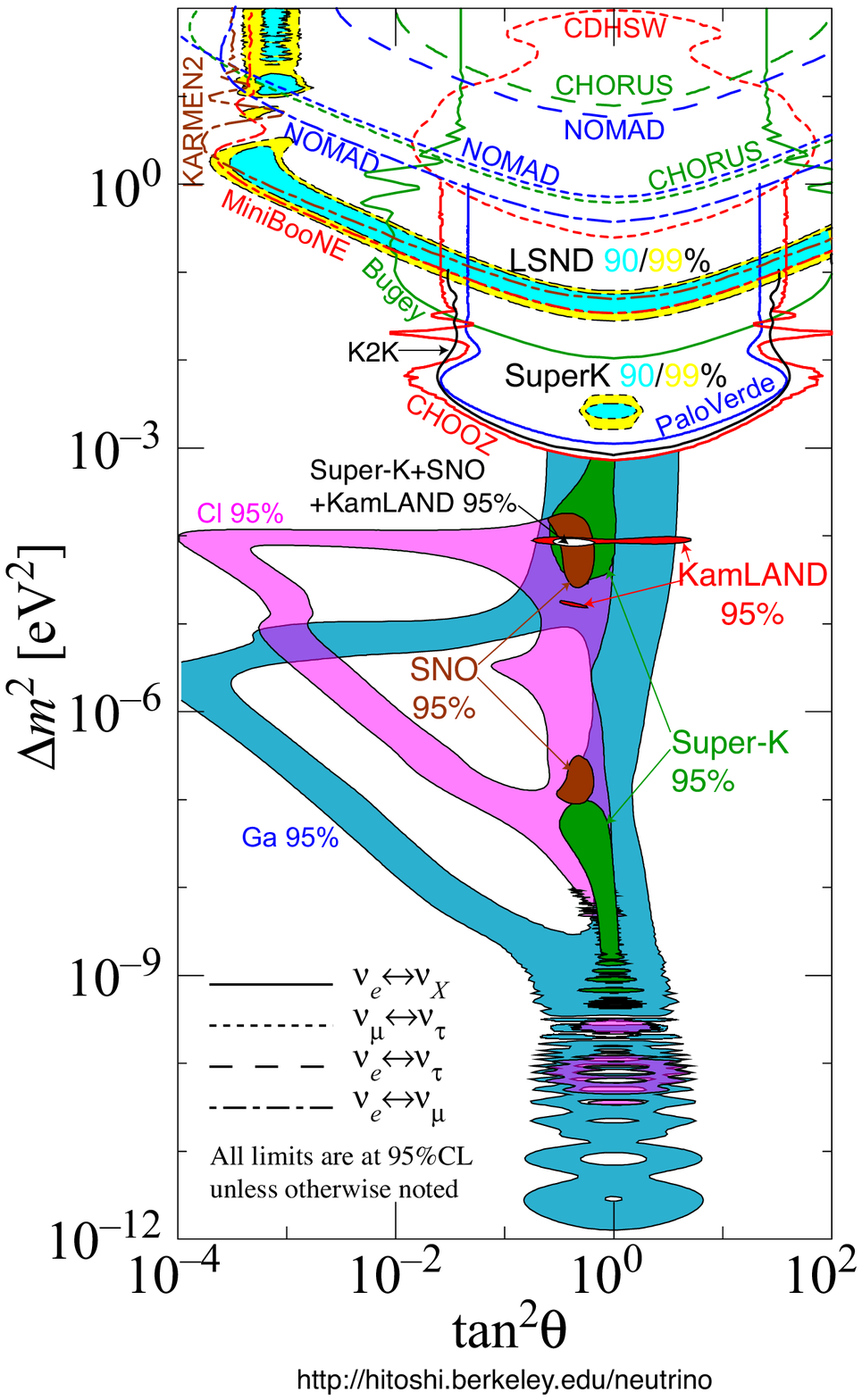}
\vspace*{-4mm}
    \caption{Summary of the currently allowed regions from a global analysis
of atmospheric and solar neutrino experiments including first results
from KamLAND (taken from H.Murayama's web site
http://hitoshi.berkeley.edu/neutrino/ .)}
\label{fig0}
\vspace*{-2mm}
\end{figure}

\begin{figure}
\includegraphics[width=0.76\textwidth]{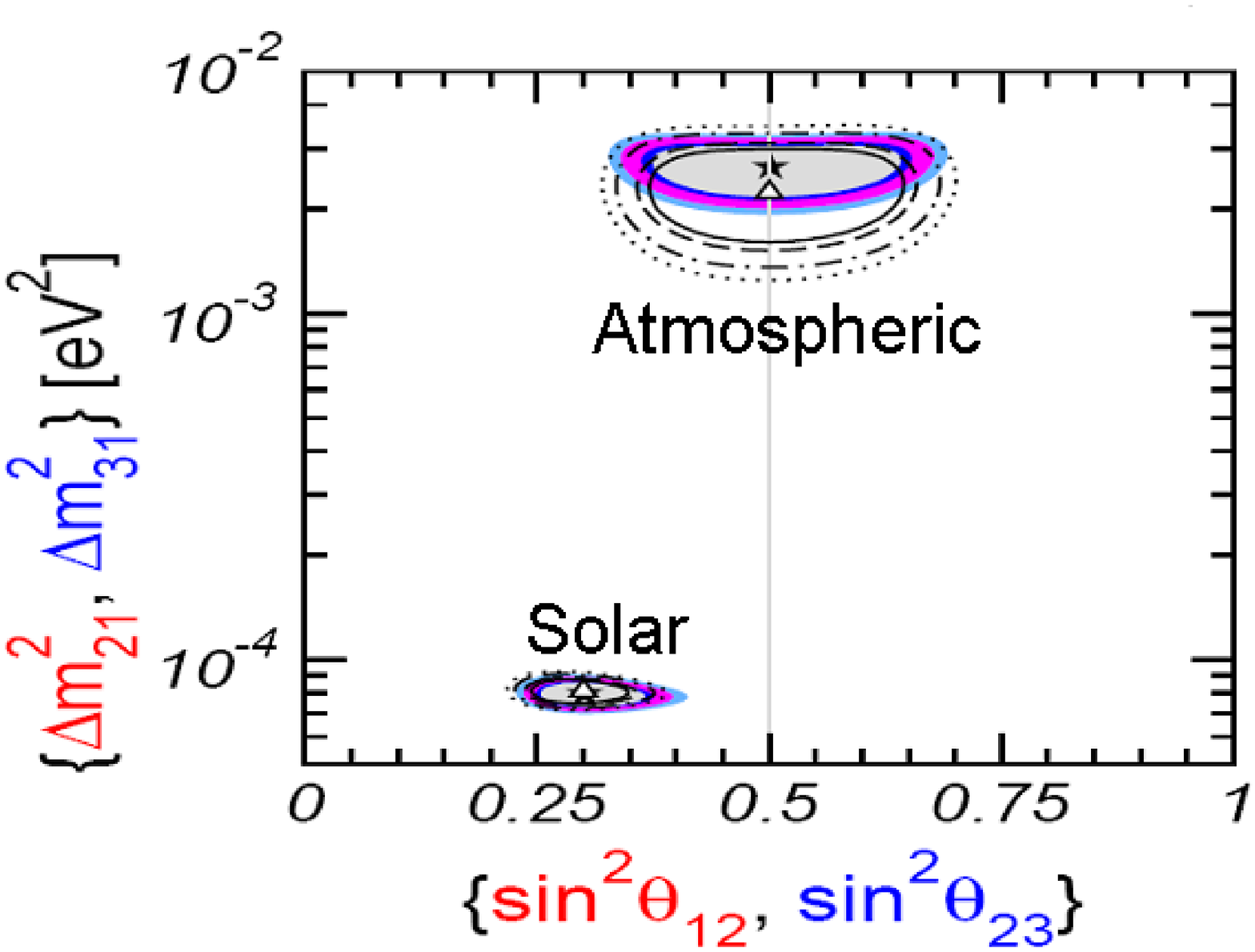}
\vspace*{-4mm}
    \caption{Summary of the currently allowed regions from a global analysis
of atmospheric and solar neutrino experiments, taken from
\cite{Maltoni:2004ei} where details concerning these plots may be found.}
\label{valle}
\vspace*{-2mm}
\end{figure}

In summary,
evidence for neutrino oscillations comes from a wide variety of
sources, and the current status of all neutrino oscillation experiments
is summarized in Fig.\ref{fig0}.
Though this figure is rather busy, the allowed atmospheric region
can be identified by its high value of
$\Delta m^2 \approx 3 \times 10^{-3}\ eV^2$, corresponding to the
region labelled ``SuperK 90/99\%'' in Fig.\ref{fig0}.
The allowed solar region
can be located from its value of
$\Delta m^2 \approx 8 \times 10^{-5}\ eV^2$, corresponding to the
intersection of the upper SNO kidney shaped region
with the thin upper KamLAND region in Fig.\ref{fig0}.
These allowed atmospheric and solar regions are plotted again
in Fig.\ref{valle}, and correspond to the
values summarized in Table \ref{tab:globalallowedrange} \cite{Maltoni:2004ei}.

\begin{table}
  \begin{center}
    \begin{tabular}{|c|c|c|c|c|}
    \hline
        parameter & best fit & 2$\sigma$ & 3$\sigma$ & 4$\sigma$
        \\
        \hline
        $\Delta m^2_{21}\: [10^{-5}$ eV$^2]$
        & 7.9 & 7.3--8.5 & 7.1--8.9 & 6.8--9.3\\
        $\Delta m^2_{31}\: [10^{-3}$ eV$^2]$
        & 2.6 & 2.2--3.0 & 2.0--3.2 & 1.8--3.5\\
        $\sin^2\theta_{12}$
        & 0.30 & 0.26--0.36 & 0.24--0.40 & 0.22--0.44\\
        $\sin^2\theta_{23}$
        & 0.50 & 0.38--0.63 & 0.34--0.68 & 0.31--0.71 \\
        $\sin^2\theta_{13}$
        & 0.000 &  $\leq$ 0.025 & $\leq$ 0.040  & $\leq$ 0.058 \\
        \hline
    \end{tabular}
    \caption{
      \label{tab:globalallowedrange}
      Best-fit values, 2$\sigma$,
      3$\sigma$, and 4$\sigma$ intervals (1 dof) for the
      three--flavour neutrino oscillation parameters from global data
      including solar, atmospheric, reactor (KamLAND and CHOOZ) and
      accelerator (K2K and MINOS) experiments, taken from \cite{Maltoni:2004ei}.
    }
  \end{center}
\end{table}

\subsection{Tri-bimaximal mixing}

The current experimental situation for neutrino mixing can be
summarized by:
$\sin^2 \theta_{23} = 0.5\pm 0.1$,
$\sin^2 \theta_{12} = 0.30\pm 0.03$, $\sin^2 \theta_{13} <0.04$.
Maximal mixing corresponds to $\sin^2 \theta_{23} = 1/2$,
and to first order in the small reactor angle the lepton mixing matrix
can then be written as:
\begin{equation}
U\approx
\left( \begin{array}{rrr}
c_{12} & s_{12} & \theta_{13}\\
-\frac{s_{12}}{\sqrt{2}} &
\frac{c_{12}}{\sqrt{2}} &
\frac{1}{\sqrt{2}}\\
\frac{s_{12}}{\sqrt{2}} &
-\frac{c_{12}}{\sqrt{2}}&
\frac{1}{\sqrt{2}}
\end{array}
\right).
\label{approxmix}
\end{equation}
Tri-bimaximal lepton mixing \cite{tribi}:
corresponds to the choice:
$\sin^2 \theta_{23} = 1/2$,
$\sin^2 \theta_{12} = 1/3$, $\sin^2 \theta_{13} =0$,
\begin{eqnarray}
U\approx
\left( \begin{array}{rrr}
\sqrt{\frac{2}{3}}  & \frac{1}{\sqrt{3}} & 0 \\
-\frac{1}{\sqrt{6}}  & \frac{1}{\sqrt{3}} & \frac{1}{\sqrt{2}} \\
\frac{1}{\sqrt{6}}  & -\frac{1}{\sqrt{3}} & \frac{1}{\sqrt{2}}
\end{array}
\right).
\label{tribi}
\end{eqnarray}
In terms of the coloured bands in Fig.\ref{fig1} tri-bimaximal
mixing corresponds to the following:
the state $\nu_3$ with mass $m_3$ consists of
a half and half mixture of $\nu_{\mu}$ and $\nu_{\tau}$;
the state $\nu_2$ with mass $m_2$ is made up of equal thirds of
$\nu_{e}$, $\nu_{\mu}$ and $\nu_{\tau}$;
and the state $\nu_1$ with mass $m_1$ comprises
two thirds $\nu_{e}$, a sixth $\nu_{\mu}$ and a sixth $\nu_{\tau}$.

Assuming tri-bimaximal mixing there is a very simple interpretation
of atmospheric and solar neutrino oscillations.
The interpretation of atmospheric oscillations is that
the muon neutrino $\nu_{\mu}$ contains a large admixture
$|U_{\mu 3}|= 1/\sqrt{2}$ of the third mass eigenstate
$\nu_3= \frac{\nu_{\mu}+\nu_{\tau}}{\sqrt{2}}$,
giving a maximal mixing of $\nu_{\mu}$ with $\nu_{\tau}$,
with an average probability of 1/2 of finding a $\nu_{\tau}$
in an initial pure $\nu_{\mu}$ beam.
The interpretation of solar oscillations is that the electron
neutrino $\nu_e$ contains a large admixture
$|U_{e2}|= 1/\sqrt{3}$ of a second mass eigenstate
$\nu_2= \frac{\nu_e + \nu_{\mu}-\nu_{\tau}}{\sqrt{3}}$, giving
trimaximal mixing of $\nu_e$ with $\nu_{\mu}$ and $\nu_{\tau}$,
with an average probability of 1/3 each of finding a $\nu_{\mu}$
or a $\nu_{\tau}$ in an initial $\nu_e$ beam,
and an average probability of 1/3 that the $\nu_e$ remains a $\nu_e$.
\footnote{The small admixture $|U_{e3}|<0.2$ of the third mass
eigenstate $\nu_3\approx \frac{\nu_{\mu}+\nu_{\tau}}{\sqrt{2}}$
does not play an important role in solar neutrino
oscillations.}.

Given the one sigma experimental errors above, there is no
good reason to believe that lepton mixing takes the tri-bimaximal
form exactly. However it clearly is consistent with the
data, at worst gives a nice mnemonic for the lepton mixing matrix,
and at best can provide some clues for the construction of a theory
of neutrino mixing.

\subsection{The LSND signal}
The signal of another independent mass splitting from the LSND
(Liquid Scintillator Neutrino Detector)
accelerator experiment \cite{Athanassopoulos:1997pv}.
The LSND collaboration found an excess of electron antineutrinos
from a beam of neutrinos consisting of the
decay products of a pion particle beam at
the Los Alamos Meson Physics Facility (LAMPF) accelerator in New Mexico.
The conclusion was that muon
antineutrinos in the beam
were changing into electron antineutrinos while propagating.
This would either
require a further light neutrino state with no weak interactions
(a so-called ``sterile neutrino'') or some other non-standard
physics. This effect has not been confirmed by a similar
experiment KARMEN \cite{Eitel:2000by}, and a decisive experiment
MiniBooNE has recently reported its first results
\cite{AguilarArevalo:2007it}. In Figure \ref{fig0} the LSND signal
region is indicated, together with the KARMEN and MiniBooNE
excluded regions. In particular MiniBooNE excludes the simplest
two neutrino oscillation interpretation of the LSND signal at 98\%
C.L. \cite{AguilarArevalo:2007it}. Indeed there seems to be no
particular motivation for including light sterile neutrinos at the
present time coming from theory, experiment or cosmology.

\subsection{Experimental Prospects and Challenges}

Neutrino physics has, now entered the precision era.
Future neutrino oscillation experiments,
will give accurate information about the mass squared splittings
$\Delta m_{ij}^2\equiv m_i^2-m_j^2$, mixing angles, and the
CP violating phase $\delta$. Long baseline neutrino beam
experiments will given more accurate determinations
of the atmospheric parameters, eventually to 10\%.

MINOS (Main Injector Neutrino Oscillation Search) experiment was also
proposed in 1995, with a neutrino beam pointed from Fermilab to the
Soudan mine in Minnesota, with a baseline of 735km.  The experiment
started running in the spring of 2005, and within a year had gathered
data corresponding to $1.27 \times 10^{20}$ protons on target.  The
first results from MINOS \cite{Michael:2006rx} were shown in
Fig.~\ref{fig:MINOS-Oscillation-Contours-With-K2K-Super-K}. MINOS will
run for 5 years, with a goal of accumulating $16 \times 10^{20}$
protons on target, which should improve our knowledge of the
oscillation parameters dramatically.
In addition a neutrino beam from
CERN to the OPERA detector in the Gran Sasso tunnel is presently
underway, and experimenters are looking for $\tau$ tracks to prove conclusively
that muon neutrinos oscillate to tau neutrinos.

In the next couple of years T2K \cite{Itow:2001ee},
a Japanese experiment sending a neutrino
beam from the J-PARC complex to Super-Kamiokande is due to start.
It will be an  ``off-axis superbeam'' over a baseline of 295 km.
Neutrino beams originate from charged pion decays, which generally results in
a large spread of neutrino energies. However, for a specific angle relative
to the pion direction, the neutrinos have a quite monochromatic
energy spectrum and therefore such ``off-axis'' neutrino beams
will have quite a well defined energy, which can be advantageous
for certain experimental measurements.
Its first goal is to measure $\theta_{13}$ or set a limit
on it of about 0.05 (as compared to the CHOOZ limit
on $\theta_{13}$ of about 0.2).
Interestingly MINOS over a LBL of 735 km is more sensitive than
J-PARC to matter effects, so there should be some interesting
complementarity between these two experiments,
which could for example allow the sign of $\Delta m^2_{32}$
to be determined. An ``off-axis superbeam'' version of the MINOS
experiment called NO$\nu$A \cite{Ayres:2004js} is seeking approval in the US.

The ultimate goal of oscillation experiments however is to measure
the CP violating phase $\delta$. To do this
it would seem that all the stops would need to be pulled out
in neutrino physics experiments. Various Superbeam,
or Beta-beam or Neutrino Factory options are currently being
considered. For example an upgraded J-PARC
with a 4MW proton driver and a 1 megaton Hyper-Kamiokande
detector \cite{Itow:2001ee}, or some sort of Neutrino Factory based on
muon storage rings would seem to be required for this purpose
\cite{Geer:1997iz}.

However oscillation experiments are not capable of telling
us anything about the absolute scale of neutrino masses.
Tritium beta decay end point experiments
measure the ``electron neutrino mass'' defined by
\beq
m_{\nu_e}\equiv \sqrt{\sum_i|U_{ei}|^2m_i^2}.
\eeq
The present Mainz limit is 2.2 eV \cite{Kraus:2004zw}.
The forthcoming KATRIN \cite{Osipowicz:2001sq}
experiment has a proposed sensitivity of
0.35 eV.

Establishing whether the neutrinos with definite mass
$\nu_j$ are Dirac fermions possessing distinct antiparticles,
or Majorana fermions, i.e., spin 1/2 particles that
are identical with their antiparticles, is
of fundamental importance for understanding
the origin of $\nu$-masses and mixing and
the underlying symmetries of particle
interactions.
The only experiments which have the
potential of establishing the
Majorana nature of massive neutrinos are the
neutrinoless double beta-decay experiments searching
for the nuclear decay process $(A,Z) \rightarrow (A,Z+2) + e^- + e^-$,
where $A$ is the number of protons plus neutrons
and $Z$ is the number of protons in the decaying nucleus
(for a review see e.g.\cite{Zuber:2004bx,Aalseth:2004hb}).
Neutrinoless double beta is only sensitive to
Majorana masses and effectively measures the combination
\beq
<m_{\beta \beta_{0\nu}}>\equiv |\sum_i|U_{ei}|^2m_ie^{i\alpha_i}| .
\eeq
Note the appearance of the Majorana phases $\alpha_i$ from the
fourth matrix in Fig.\ref{MNS2}. These phases can lead to
cancellations in the sum over the mass flavours, where a
precise cancellation would correspond to a Dirac mass.
Such experiments are very important since they would not
only establish the neutrino mass scale, but would also
establish the nature of the neutrino mass, since the process
is only allowed if neutrinos have Majorana mass.
There has been a recent claim of a signal in neutrinoless double
beta decay correponding to $<m_{\beta \beta_{0\nu}}>\approx 0.4$ eV
in an analysis of the Heidelberg-Moscow $^{76}Ge$ experiment
\cite{Klapdor-Kleingrothaus:2001ke}.
However this claim has been criticised by two groups
\cite{Feruglio:2002af}, \cite{Aalseth:2002dt} and in turn this
criticism has been refuted \cite{Klapdor-Kleingrothaus:2002kf},
followed by a further paper containing
a more refined analysis \cite{KlapdorKleingrothaus:2004wj}.
This claim will be directly tested in the near future by other forthcoming
$^{76}Ge$ experiments such as Majorana and GERDA which will
achieve sensitivies of about $<m_{\beta \beta_{0\nu}}>\approx
0.05-0.1$ eV \cite{Aalseth:2004hb}.
Similar sensitivities are also planned in different isotopes
by other forthcoming experiments such as CUORE, Super-NEMO,
COBRA, EXO \cite{Aalseth:2004hb} or SNO++, the recent exciting proposal
to fill the now decomissioned SNO vessel with liquid
scintillator doped with Neodymium \cite{SNO++}.
The most ambitious sensitivities planned are down to $<m_{\beta
  \beta_{0\nu}}>\approx 0.01 $ eV \cite{Aalseth:2004hb}.

Let us end this section by
summarizing the main experimental challenges facing neutrino physics
at the present time. The following challenges
can be addressed by future neutrino oscillation experiments:
\begin{itemize}
  \item
    The sign of $\Delta m^2_{31}$: whether the neutrino mass
    ordering is ``normal'' or ``inverted'' has important implications for
    Grand Unification, Flavour Models and Cosmology.
  \item
    The question of CP-violation ($\delta$): measurement of
    the oscillation phase represents
    the Holy Grail of neutrino physics, since it would signal
    CP violation in the lepton sector, which would also have profound
    implications for Grand Unification, Flavour Models and Cosmology.
  \item
    High precision measurements of mixing angles: especially
    $\theta_{13}$ which has so far not been measured at all;
    a high precision determination of all the mixing angles
    again provides crucial information for Grand Unification and Flavour Models.
\end{itemize}
The remaining challenges
can be addressed by a combination of Neutrinoless Double Beta
decay experiments, Tritium end-point experiments and Cosmological
considerations:
\begin{itemize}
  \item
    Majorana vs Dirac: the question of whether neutrino masses
    are Majorana or Dirac in nature has profound implications
    for particle physics.
  \item
    The absolute neutrino mass scale: only mass squared
    differences are relevant for neutrino oscillations, and the
    absolute neutrino mass scale is so far not measured.
\end{itemize}

\section{Implications for particle physics and cosmology}
\label{implications}

In this section we discuss the origin and nature of neutrino mass,
and emphasize that, whatever its origin, it must correspond
to new physics Beyond the Standard Model.
We then discuss the see-saw mechanism, which is
a natural and appealing explanation of small neutrino
masses, and its application to atmospheric and solar
oscillation data using the sequential dominance mechanism.
We also discuss an alternative explanation of small
neutrino masses in terms of extra space dimensions.
Finally we discuss some cosmological implications of
neutrino mass.

\subsection{The origin of neutrino mass}
Neutrino mass is
zero in the Standard Model for three independent reasons:
\begin{enumerate}
\item There are no right-handed neutrinos $\nu_R$
\item There are only Higgs doublets $(H^+, H^0)$
\item The theory is renormalizable
\end{enumerate}
In the SM these conditions all apply and so neutrinos
are massless with $\nu_e$, $\nu_{\mu}$, $\nu_{\tau}$
distinguished by separate lepton numbers
$L_e$, $L_{\mu}$, $L_{\tau}$. Neutrinos and antineutrinos
are distinguished by total conserved lepton number
$L=L_e+L_{\mu}+L_{\tau}$. To generate neutrino mass we must relax
one or more of these conditions.
For example, by adding
right-handed neutrinos
the Higgs mechanism of the Standard Model can give neutrinos
the same type of mass as the electron mass or other
charged lepton and quark masses.

\begin{figure}[t]
\centering
\includegraphics[width=0.76\textwidth]{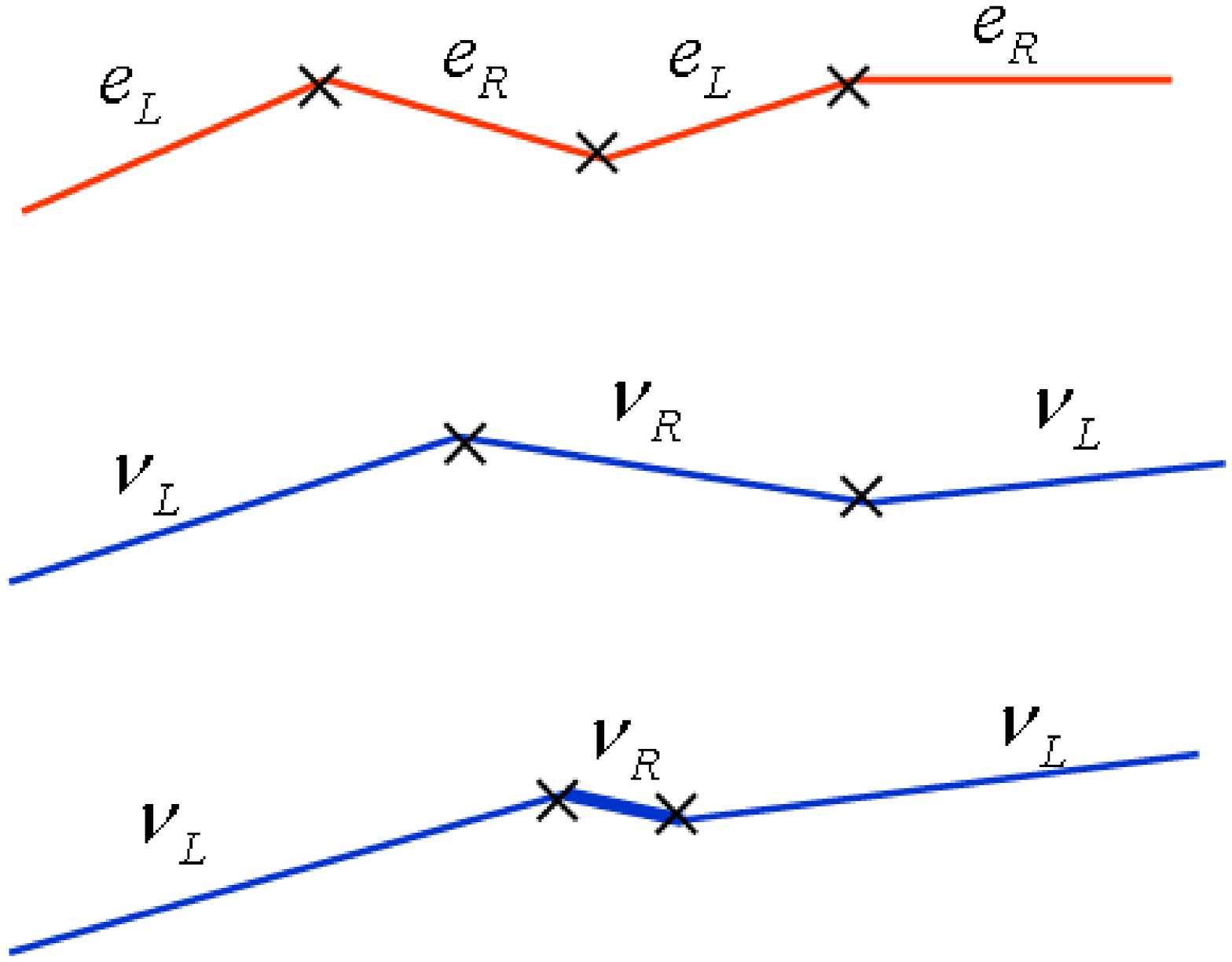}
\vspace*{-4mm}
    \caption{\footnotesize A diagramatic illustration of fermion
masses in the presence of a background Higgs field $H^0$
which is uniformly switched on in the vacuum. In the upper
diagram, a left-handed electron mass interacts with the
background Higgs field to become a right-handed electron, then
interacts again to become a left-handed electron, and so on,
resulting in a Dirac mass $m_e$ for the electron.
In the centre diagram a similar thing can happen to the
neutrino provided right-handed neutrinos are introduced
into the Standard Model, leading to a Dirac neutrino mass $m_{LR}$.
The lower diagram shows what happens when the right-handed neutrino
acquires a large mass $M_{RR}$ independently of the Higgs
mechanism. In this case, the heavy right-handed neutrino
cannot travel very far due to its large mass,
and in the limit of extremely large $M_{RR}$,  when the length of
its propagation goes to zero, the lower diagram looks effectively like
a direct interaction between two left-handed neutrinos, resulting
in an effective left-handed Majorana mass
$m^{eff}_{LL}=m^{2}_{LR}/M_{RR}$.
This is called the see-saw mechanism.} \label{murayama2}
\vspace*{-2mm}
\end{figure}

\begin{figure}[t]
\centering
\includegraphics[width=0.76\textwidth]{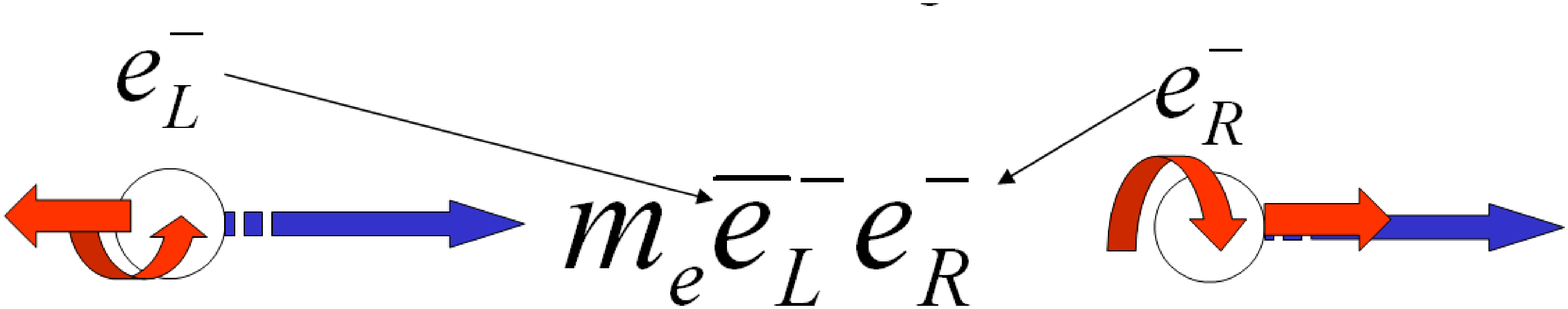}
\vspace*{-4mm}
    \caption{\footnotesize The electron Dirac mass $m_e$ can
be thought of as an interaction between a left-handed
electron $e_L^-$ and a right-handed electron $e_R^-$.
In these figures the long
(blue) arrows denote the electron momentum vector, and the short
(red) arrows denote the electron spin vector. For right-handed
electrons $e_R^-$ the spin vector and the momentun vector are aligned,
whereas for left-handed electrons $e_L^-$ they are anti-aligned.
Such mass terms $m_e\bar{e}^-_L e^-_R$
appear in the Lagrangian
density for the quantum field theory, where the
bar over the $e_L^-$ has a conventional meaning
that need not concern us here. From our point
of view here such mass terms may simply be regarded as interactions that
enable left-handed electrons to interact with
right-handed electrons.
} \label{mass1}
\vspace*{-2mm}
\end{figure}

We begin by discussing the Higgs mechanism of the Standard Model.
The Higgs mechanism, originally proposed by the British physicist Peter Higgs,
is the mechanism that gives mass to all elementary particles in particle physics.
It makes the W boson different from the photon, for example.
It can be understood as an elementary case of tachyon condensation
where the role of the tachyon is played by a scalar field called the Higgs field.
The massive quantum excitation of the Higgs field is also called the Higgs boson.
According to the Standard Model all of space is filled by
a background Higgs field, which is somewhat analagous to
the background electric and magnetic fields that
are also present in deep space. In the Standard Model
the background Higgs field is due to a single doublet
consisting one charged and one neutral Higgs field
$(H^+, H^0)$, where only the neutral field $H^0$ is switched on
in the vacuum, breaking the symmetry of the doublet,
and hence breaking the symmetry between the weak and the
electromagnetic interactions, resulting in $W,Z$ masses.
It also results in fermion masses due to their interaction
with the background Higgs field. As an electron travels through
space it is continually interacting with the background Higgs field
as illustrated in the upper diagram in Fig.\ref{murayama2},
resulting in its mass. However, with each interaction its
handedness changes, so that its mass can be thought of as
an interaction between a left-handed
electron $e^-_L$ and a right-handed electron $e^-_R$
as shown in Fig.\ref{mass1}.
Such an interaction
gives rise to what is known as a Dirac mass, named after Paul Dirac,
an English physicist who proposed the equation describing massive
electrons that bears his name. Strictly speaking such mass
terms appear in the Lagrangian
density for the quantum field theory, but from our point
of view here they may simply be regarded as interactions
between a left-handed electron and a right-handed electron,
and no knowledge of quantum field theory is required
to understand this basic point.

\begin{figure}[t]
\centering
\includegraphics[width=0.76\textwidth]{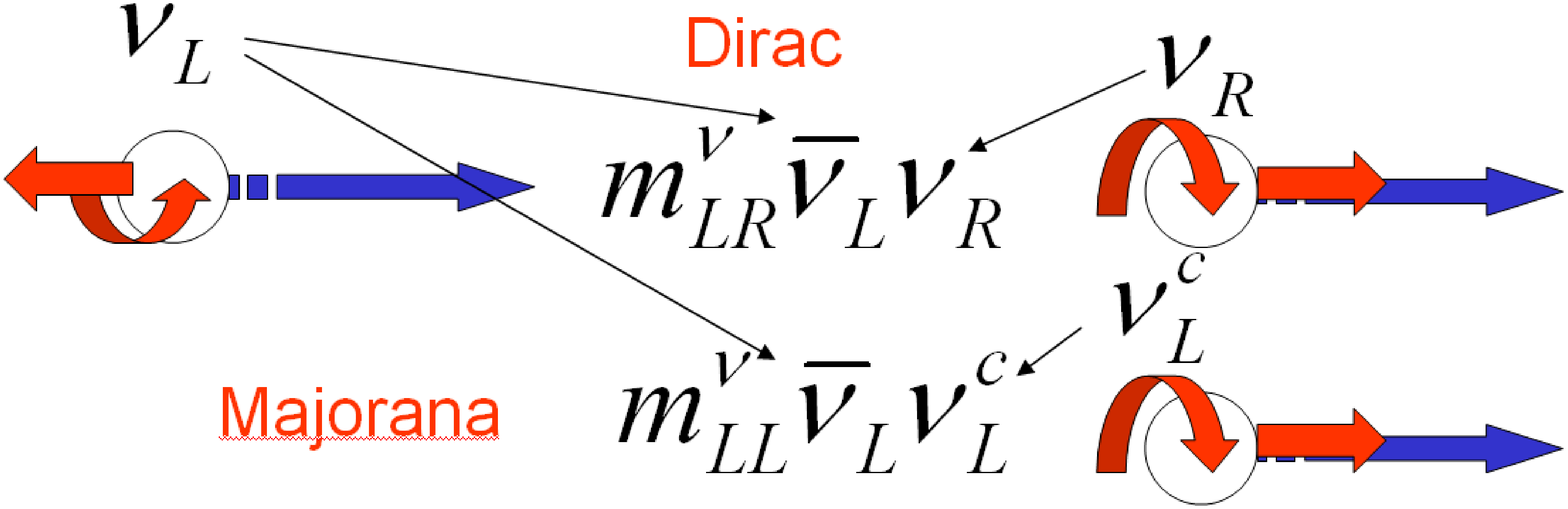}
\vspace*{-4mm}
    \caption{\footnotesize
For neutrinos there are two types of mass that
are possible. As in the case of the electron there is the Dirac mass
$m^{\nu}_{LR}$
that couples a left-handed neutrino $\nu_L$ to a right-handed
neutrino $\nu_R$, as shown in the upper part of the diagram.
However, the role of a right-handed neutrino can be played
by $\nu_L^c$ obtained by transforming the left-handed neutrino
$\nu_L$ under the operations charge and parity conjugatation,
where $\nu_L^c$ is a right-handed antineutrino.
If $\nu_L$ interacts with $\nu_L^c$ then this results in
a Majorana mass $m^{\nu}_{LL}$.
Such mass terms
appear in the Lagrangian
density for the quantum field theory, where the
bar over the $\nu_L$ has a conventional meaning
that need not concern us here. From our point
of view here such mass terms may simply be regarded as interactions.
} \label{mass2}
\vspace*{-2mm}
\end{figure}

\begin{figure}[t]
\centering
\includegraphics[width=0.76\textwidth]{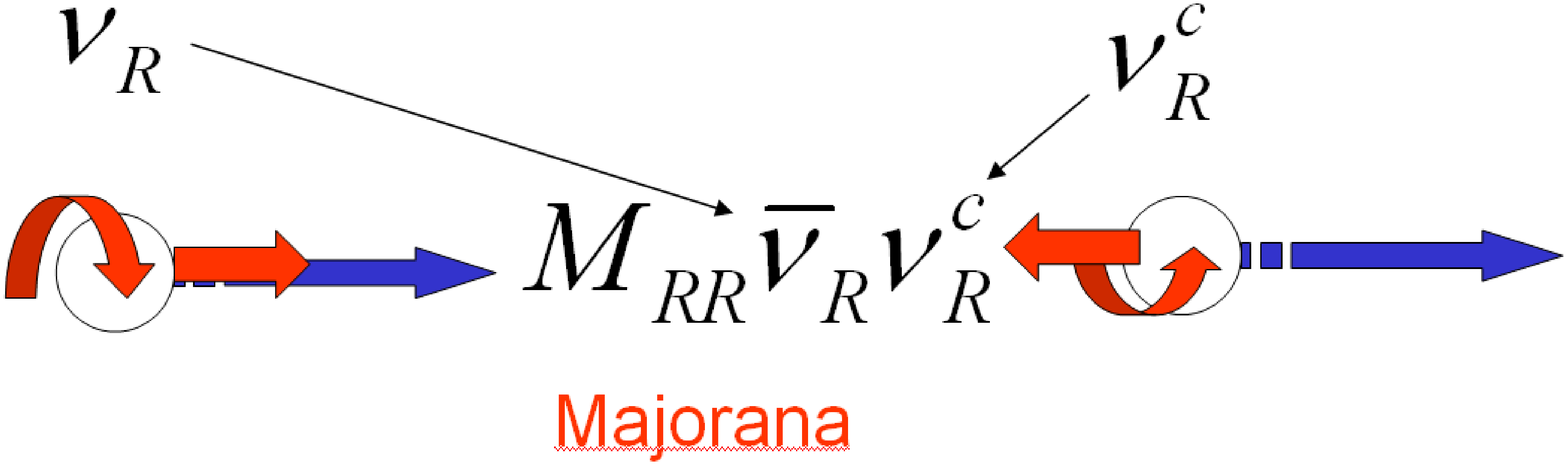}
\vspace*{-4mm}
    \caption{\footnotesize
If right-handed neutrinos $\nu_R$ are added to the
Standard Model, then they can also acquire a Majorana mass $M_{RR}$
by coupling to their own charge and parity conjugated
states $\nu_R^c$. Since it does not take part in the Standard Model weak
interactions, its Majorana mass $M_{RR}$ may be consistently taken
to be much larger than the $W,Z$ boson masses,
and may be arbitrarily large.
Such mass terms
appear in the Lagrangian
density for the quantum field theory, where the
bar over the $\nu_R$ has a conventional meaning
that need not concern us here. From our point
of view here such mass terms may simply be regarded as interactions that
enable right-handed neutrinos to interact with
left-handed antineutrinos.
 } \label{mass3}
\vspace*{-2mm}
\end{figure}

It is possible to add right-handed neutrinos $\nu_R$ to the Standard
Model, providing that the right-handed neutrinos do not take part
in the weak interaction so as to not contradict with the result
of Goldhaber et al that weakly interacting
neutrinos are always left-handed.
With right-handed neutrinos present a similar interaction can take
place as for electrons, giving rise to a Dirac mass
for the neutrino $m_{LR}$, as shown in the centre diagram
in Fig.\ref{murayama2} and the upper part of the diagram
in Fig.\ref{mass2}. In principle it is also possible to give neutrinos
a new kind of mass called a Majorana mass $m_{LL}$, named after
the Sicilian physicist, Ettore Majorana, if the left-handed
neutrino $\nu_L$ interacts
with its own charge and parity conjugated state, the
right-handed antineutrino $\nu_L^c$, where the superscript $c$ denotes
the simultaneous operation of charge conjugation (C) (replacing the
particle by the antiparticle)
and parity (P) (replacing the particle by its mirror image, which has the
effect of reversing the spin direction).
Such a Majorana mass $m_{LL}$ is shown in the lower part of Fig.\ref{mass2}.
In principle
right-handed neutrinos $\nu_R$ can also independently acquire their
own Majorana masses $M_{RR}$, by interacting with their own CP conjugates
$\nu_R^c$ as shown in Fig.\ref{mass3}.
Such Majorana masses $m_{LL}$ or $M_{RR}$ are only possible in principle for
neutrinos since they are the only leptons which are electrically neutral.
If such existed, however, they would violate total lepton
number $L$.

Although left-handed Majorana masses $m_{LL}$ are possible
in principle, in the Standard Model they are zero since
the background Higgs field $H^0$ is incapable of flipping a $\nu_L$
into a $\nu_L^c$. If the background Higgs field $H^0$ were a component
of a Higgs triplet $(H^{++},H^+, H^0)$ instead of a Higgs doublet
$(H^+, H^0)$ then such a flipping would be possible.
However in the Standard Model only Higgs doublets are present
and then $H^0$ can only flip a $\nu_L$ into a $\nu_R$,
as seen in Fig.\ref{murayama2}.
However there is nothing to prevent
the right-handed neutrinos $\nu_R$ having
Majorana masses $M_{RR}$, where the magnitude of such masses can
take any value, and in particular such masses could be very large.
The Heisenberg Uncertainty Principle, which allows energy conservation
to be violated on small time intervals,
then allows a left-handed neutrino to convert into a
heavy right handed neutrino, via the Higgs interaction,
for a brief moment before reverting back to
being a left-handed neutrino, as shown in
the lower diagram in Fig.\ref{murayama2}.
For a very large $M_{RR}$, this effectively results in a very small
effective Majorana mass for the left-handed neutrino,
$m_{LL}^{eff}=(m^{\nu}_{LR})^2/M_{RR}$.
The presence of large right-handed Majorana masses $M_{RR}$
therefore leads to
an attractive mechanism for explaining the smallness of
neutrino masses compared to charged fermion masses.
This is the so-called see-saw mechanism.
The smallness of the neutrino mass $m_{LL}^{eff}$
is associated with the heaviness of the right-handed neutrino
mass $M_{RR}$.

The third requirement for the absence of neutrino mass in
the Standard Model is that the theory is renormalizable.
This is a technical requirement that all the interactions
of the theory are generated by particle exchange,
and that quantum corrections to the theory do not introduce any
infinities. A simple example of a non-renormalizable
interaction that would generate neutrino mass
would be a ``contact interaction'' between two
left-handed neutrinos and two Higgs fields,
corresponding to the lower diagram in Fig.\ref{murayama2}
but with the right-handed neutrino line shrunk to zero.
In this case a Majorana mass
$m_{LL}^{eff}=(m^{\nu}_{LR})^2/\Lambda $ would
be generated but it would not be due to
the exchange of heavy right-handed neutrinos but due to
the non-renormalizable ``contact interaction'' where the
Standard Model is valid up to some cut-off $\Lambda$.
In fact the lower diagram in Fig.\ref{murayama2}, with
very heavy right-handed neutrinos, is well
approximated by such a non-renormalizable ``contact interaction'',
since the $M_{RR}$ is very large and its propagation
length is very small, so in this case we would identify
$\Lambda = M_{RR}$. Thus, even if non-renormalizable ``contact
interactions'' are added to the Standard Model they may be due to
very heavy particle exchange. Of course the origin of the
contact interaction may be due to the exchange of other particles
of mass $\Lambda$, different from heavy right-handed neutrinos.

\subsection{The see-saw mechanism}
In this subsection we discuss the see-saw mechanism a little
more quantitatively.
Let us first summarize
the different types of neutrino mass that are possible.
There are Majorana masses of the form
\beq
m_{LL}\overline{\nu_L}\nu_L^c
\label{mLL}
\eeq
where $\nu_L$ is a left-handed neutrino field and $\nu_L^c$ is
the CP conjugate of a left-handed neutrino field, in other words
a right-handed antineutrino field.
Such mass terms have been discussed in the previous section,
and have been represented diagrammatically in Fig.\ref{mass2}.
Strictly speaking such mass terms appear in the Lagrangian
density for the quantum field theory, but from our point
of view here they may simply be regarded as interactions that
enable left-handed neutrinos to interact with
right-handed antineutrinos, as depicted in Fig.\ref{mass2}.

Such Majorana masses are possible
to since both the neutrino and the antineutrino
are electrically neutral and so
Majorana masses are not forbidden by electric charge conservation.
For this reason a Majorana mass for the electron would
be strictly forbidden. However such Majorana neutrino masses
violate lepton number conservation, and in the standard model,
assuming only the simplest Higgs bosons are present, are
forbidden.
The idea of the simplest version of the see-saw mechanism is to assume
that such terms are zero to begin with, but are generated effectively,
after right-handed neutrinos are introduced \cite{seesaw}.

If we introduce right-handed neutrino fields then there are two sorts
of additional neutrino mass terms that are possible. There are
additional Majorana masses of the form
\beq
M_{RR}\overline{\nu_R}\nu_R^c
\label{MRR}
\eeq
where $\nu_R$ is a right-handed neutrino field and $\nu_R^c$ is
the CP conjugate of a right-handed neutrino field, in other words
a left-handed antineutrino field. In addition there are
Dirac masses of the form
\beq
m_{LR}\overline{\nu_L}\nu_R.
\label{mLR}
\eeq
Such Dirac mass terms conserve lepton number, and are not forbidden
by electric charge conservation even for the charged leptons and
quarks.

The Higgs mechanism, in its simplest form at least,
forbids Majorana masses of the type $m^{\nu}_{LL}$,
involving the left-handed neutrino $\nu_L$, and its CP conjugate $\nu_L^c$,
but permits Majorana masses $M_{RR}$ involving purely right-handed neutrinos
$\nu_R$ and its CP conjugate $\nu_R^c$.
In fact just as $m^{\nu}_{LL}$ must be zero in the Standard Model,
so $M_{RR}$ may be arbitarily large.
The reason is essentially that the left-handed neutrino $\nu_L$
takes part in weak interactions with the $W,Z$ bosons,
and if it were very heavy it would disturb the theory.
The right-handed neutrino $\nu_R$ on the other hand
does not take part in weak interactions with the $W,Z$ bosons,
and so its mass $M_{RR}$ can be arbitarily large.

With the types of neutrino mass discussed
in Eqs.\ref{MRR},\ref{mLR} (but not Eq.\ref{mLL} since we
assume no Higgs triplets) we have the see-saw mass matrix
\begin{equation}
\left(\begin{array}{cc} \overline{\nu_L} & \overline{\nu^c_R}
\end{array} \\ \right)
\left(\begin{array}{cc}
0 & m_{LR}\\
m_{LR}^T & M_{RR} \\
\end{array}\right)
\left(\begin{array}{c} \nu_L^c \\ \nu_R \end{array} \\ \right)
\label{matrix}
\end{equation}
Since the right-handed neutrinos are electroweak singlets
the Majorana masses of the right-handed neutrinos $M_{RR}$
may be orders of magnitude larger than the electroweak
scale. In the approximation that $M_{RR}\gg m_{LR}$
the matrix in Eq.\ref{matrix} may be diagonalised to
yield effective Majorana masses of the type in Eq.\ref{mLL},
\beq
m_{LL}=m_{LR}M_{RR}^{-1}m_{LR}^T.
\label{seesaw}
\eeq
The effective left-handed Majorana masses $m_{LL}$ are naturally
suppressed by the heavy scale $M_{RR}$.
In a one family example if we take $m_{LR}=M_W=80$ GeV
and $M_{RR}=M_{GUT}=10^{16}$ GeV
then we find $m_{LL}\sim 10^{-3}$ eV which looks good for solar
neutrinos.
Atmospheric neutrino masses would require
a right-handed neutrino with a mass below the GUT scale.

With three families of left-handed neutrinos and
three right-handed neutrinos the Dirac masses $m_{LR}$
are a $3\times 3$ (complex) matrix and the heavy Majorana masses $M_{RR}$
form a separate $3\times 3$ (complex symmetric) matrix.
The light effective Majorana
masses $m_{LL}$ are also a $3\times 3$ (complex symmetric) matrix and
continue to be given from Eq.\ref{seesaw} which
is now interpreted as a matrix product. From a model building
perspective the fundamental parameters which must be input
into the see-saw mechanism are the Dirac mass matrix $m_{LR}$ and
the heavy right-handed neutrino Majorana mass matrix $M_{RR}$.
The light effective left-handed Majorana mass
matrix $m_{LL}$ arises as an output according to the
see-saw formula in Eq.\ref{seesaw}.

\subsection{Sequential dominance}
We wish to apply the see-saw mechanism to account
for the atmospheric and solar mixing.
A simple and natural way to achieve a neutrino mass hierarchy
with large atmospheric and solar
mixing angles is the idea of sequential dominance
\cite{Antusch:2004gf}.
\footnote{Note that sequential dominance also works if neutrinos
are quasi-degenerate, providing one invokes
an extended type II see-saw mechanism in which
the zero in the see-saw matrix in Eq.\ref{matrix} is replaced
by a universal mass proportional to the unit matrix
\cite{Antusch:2004gf}.}

\subsubsection{Two state atmospheric mixing}
It is instructive to begin by discussing
a simple $2\times 2$ example, describing the two state
atmospheric mixing in subsection \ref{twostate}.
The starting point is to
assume that the right-handed Majorana mass
matrix and the charged lepton mass matrix are diagonal
\footnote{This is convenient since it means that one can identify the
neutrino mixing angles with the mixing angles
in Eq.\ref{euler} (recall that the lepton mixing matrix
$U$ in Eq.\ref{MNS0} involves the neutrino flavour eigenstates
which share the same electroweak doublet as the charged
lepton mass eigenstates, as discussed.)
However neither of these assumptions is
essential \cite{King:2006hn}.},
but the Dirac neutrino mass matrix is general, and then we write:
\begin{equation}
M_{RR}=
\left( \begin{array}{cc}
Y & 0     \\
0 & X
\end{array}
\right), \ \ \ \
m_{LR}=
\left( \begin{array}{cc}
e & b \\
f & c
\end{array}
\right)
\label{2by2dom}
\end{equation}
where the Dirac mass matrix elements $m_{LR}$ couple a particular
Standard Model left-handed neutrino states
$(\nu_{\mu L}, \nu_{\tau  L})$,
to a particular right-handed neutrino states
$(\nu^Y_{R }, \nu^X_{R })$ labelled by its
Majorana masses $Y,X$, respectively.
For example the element $b$ of $m_{LR}$ corresponds to
a Dirac mass coupling $\nu_{\mu L}$ to $\nu^X_{R }$.
The see-saw formula in Eq.\ref{seesaw}
$m_{LL}=m_{LR}M_{RR}^{-1}m_{LR}^T$ gives
the effective left-handed Majorana mass matrix:
\beq
m_{LL}
=
\left( \begin{array}{cc}
\frac{e^2}{Y}+\frac{b^2}{X}
& \frac{ef}{Y}+\frac{bc}{X} \\
\frac{ef}{Y}+\frac{bc}{X}
& \frac{f^2}{Y}+\frac{c^2}{X}
\end{array}
\right)
\approx
\left( \begin{array}{cc}
\frac{e^2}{Y}
& \frac{ef}{Y} \\
\frac{ef}{Y}
& \frac{f^2}{Y}
\end{array}
\right)
\label{one}
\eeq
where the approximation in Eq.\ref{one}
assumes that the right-handed neutrino of mass $Y$
is sufficiently light that it dominates in the see-saw mechanism
\cite{King:1998jw}:
\beq
\frac{e^2,f^2,ef}{Y}\gg
\frac{b^2,c^2,bc}{X}.
\label{srhndp}
\eeq
The left-handed Majorana mass matrix in Eq.\ref{one}
is now interpreted as a matrix of Majorana masses involving
the Standard Model left-handed neutrino states
$(\nu_{\mu L}, \nu_{\tau  L})$ coupling to themselves.
The physical neutrino masses $m_i$ corresponding to the
mass eigenstates $\nu_i$
are obtained by diagonalizing the
mass matrix in Eq.\ref{one},
\begin{equation}
\left(\begin{array}{cc}
 c_{23} & -s_{23} \\
 s_{23} & c_{23} \\
\end{array}\right)
\left( \begin{array}{cc}
\frac{e^2}{Y}
& \frac{ef}{Y} \\
\frac{ef}{Y}
& \frac{f^2}{Y}
\end{array}
\right)
\left(\begin{array}{cc}
 c_{23} & s_{23} \\
 -s_{23} & c_{23} \\
\end{array}\right)
=
\left( \begin{array}{cc}
0 & 0 \\
0 & \frac{e^2+f^2}{Y}
\end{array}
\right).
\label{diagone}
\end{equation}
The neutrino mass
spectrum from Eq.\ref{diagone}
then consists of one neutrino with mass $m_3\approx (e^2+f^2)/Y$
and one naturally light neutrino $m_2\ll m_3$,
since the determinant of Eq.\ref{one} is clearly
approximately vanishing, due to the dominance assumption.
The atmospheric angle from Eq.\ref{diagone} is
$\tan \theta_{23} \approx e/f$
which can be large or maximal providing $e \approx f$.
Thus two crucial features, namely a neutrino mass hierarchy
$m_3^2\gg m_2^2$ and a
large neutrino mixing angle $\tan \theta_{23} \approx 1$,
can arise naturally from the
see-saw mechanism assuming the dominance of a single right-handed
neutrino \cite{King:1998jw}.

\subsubsection{Three family neutrino mixing}
In order to account for the solar mixing angle,
we must generalise the above discussion to the
$3\times 3$ case. The generalization of Eq.\ref{2by2dom} is:
\begin{equation}
M_{RR}=
\left( \begin{array}{ccc}
Y & 0 & 0    \\
0 & X & 0 \\
0 & 0 & X'
\end{array}
\right), \ \
m_{LR}=
\left( \begin{array}{ccc}
d & a & a'    \\
e & b & b'\\
f & c & c'
\end{array}
\right).
\label{3by3dom}
\end{equation}
The generalization of Eq.\ref{srhndp} is called sequential dominance
\cite{King:1999mb}:
\beq
\frac{e^2,f^2,ef}{Y}\gg
\frac{a^2,b^2,c^2,ab,ac,bc}{X}\gg
\frac{{a'}^2,{b'}^2,{c'}^2,a'b',a'c',b'c'}{X}
\label{srhnd}
\eeq
where we also assume $d\ll e,f$.
Ignoring the small primed terms, the see-saw formula
Eq.\ref{seesaw} now gives a Majorana mass matrix in the
basis $(\nu_{e L}, \nu_{\mu L}, \nu_{\tau  L})$ as:
\beq
m_{LL}
\approx
\left( \begin{array}{ccc}
\frac{a^2}{X}+\frac{d^2}{Y}
& \frac{ab}{X}+ \frac{de}{Y}
& \frac{ac}{X}+\frac{df}{Y}    \\
\frac{ab}{X}+ \frac{de}{Y}
& \frac{b^2}{X}+\frac{e^2}{Y}
& \frac{bc}{X}+\frac{ef}{Y}    \\
\frac{ac}{X}+\frac{df}{Y}
& \frac{bc}{X}+\frac{ef}{Y}
& \frac{c^2}{X}+\frac{f^2}{Y}
\end{array}
\right)
\label{mLL2}
\eeq
Note that the lower $2\times 2$ block of Eq.\ref{mLL2} can be identified
with Eq.\ref{one}, so we expect have large atmospheric mixing as before.
The physical neutrino masses are obtained by diagonalizing the
mass matrix in Eq.\ref{mLL2}, which, ignoring phases, corresponds
to performing the Euler rotation in Fig.\ref{fig2} to go from the
basis $(\nu_{e }, \nu_{\mu }, \nu_{\tau  })$ to the
basis $(\nu_{1}, \nu_{2}, \nu_{3})$, given by
a generalization of
Eq.\ref{diagone}:
\be
U^Tm_{LL}U=\left( \begin{array}{ccc}
m_1 & 0 & 0    \\
0 & m_2 & 0 \\
0 & 0 & m_3
\end{array}
\right)
\ee
where, from Eq.\ref{euler}, $U=R_{23}R_{13}R_{12}$.
After some algebra, in a small angle $\theta_{13}$ approximation, one
finds \cite{King:1999mb}
a full neutrino mass hierarchy
\beq
m_3\gg m_2\gg m_1
\eeq
with
\be
m_3\approx \frac{e^2+f^2}{Y}, \ \ m_2\approx \frac{a^2}{Xs_{12}^2}.
\ee
Note that sequential dominance can only account for a normal
neutrino mass hierarchy, and not the inverted mass pattern.
Assuming that $d$ is negligible, the angles are determined to be
\cite{King:1999mb}:
\be
\tan \theta_{23} \approx \frac{e}{f}, \ \
\tan \theta_{12} \approx \frac{a}{bc_{23}-cs_{23}}, \ \
\tan \theta_{13} \approx \frac{a(bs_{23}+cc_{23})}{m_3X}.
\ee
Note that the solar mass and solar angle only depends
on the sub-dominant couplings.
In general large solar mixing can result. In particular
tri-bimaximal mixing in Eq.\ref{tribi} results from the
choice $e=f$, $a=b=-c$ which is called constrained sequential
dominance \cite{King:2005bj}.

\subsection{Large extra dimensions}

\begin{figure}[t]
\centering
\includegraphics[width=0.96\textwidth]{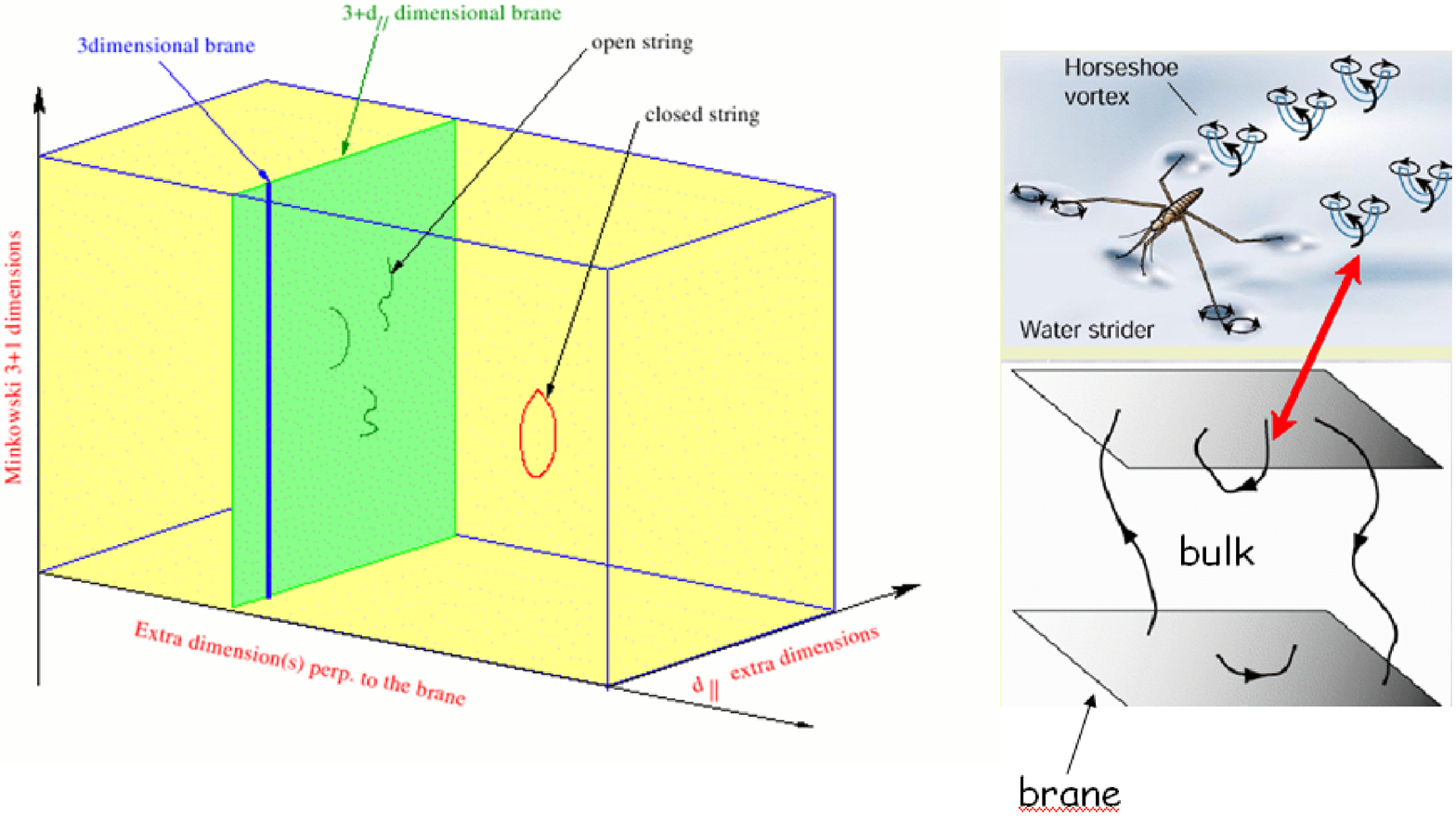}
\vspace*{-4mm}
    \caption{According to string theory there may be extra dimensions
in addition to the 3+1 of the Standard Model. According to the latest
theories, the Standard Model may reside on a 3 space dimensional
brane. Open strings attach themselves to such branes, while
closed strings float freely in the bulk of all the extra dimensions,
as indicated in the left-hand panel taken
from \cite{Antoniadis:2005aq}.
Thus we may appear as this water strider
in the right-handed panel (taken from a talk
by Jean Orloff) walking over the
membrane of the water on a pond, unaware of the bulk of water
below us.
} \label{extra}
\vspace*{-2mm}
\end{figure}

An alternative explanation of small neutrino masses comes from the
concept of extra dimensions beyond the three that we
know of, motivated by theoretical attempts to extend the Standard Model
to include gravity \cite{Antoniadis:2005aq}.
According to string theory, there may be six extra
space dimensions in addition to the three space and one time dimension
of the Standard Model. Indeed it has been suggested that the Standard
Model lives on a 3 space dimensional brane, and that we are analagous
to a bug walking on the surface of a pond, supported by the
membrane of the water, as shown in Fig.\ref{extra}.
The extra dimensions are ``compactified'' (rolled up) on circles
of small radius $R$ so that they are not normally observable.
Such extra dimensions if uniformly compactified are
called ``flat'' or if the compactification involves
a distortion or warping are called ``warped''.
It has been
suggested that right-handed neutrinos (but not the rest of the Standard
Model particles) experience one or more of these extra dimensions. The
right handed neutrinos then only spend part of their time in our world,
leading to very small Dirac neutrino masses
\cite{Arkani-Hamed:1998vp}.
In such theories there is a relation between the usual four dimensional Planck mass
$M_{Planck}\sim 10^{19}\ GeV/c^2$, the string scale $M_{string}$ and the compactification
radius of the ``flat'' extra dimensions $R$ given by:
\be
M_{Planck}^2=M_{string}^{2+n}R^n
\label{flat}
\ee
where there are $n$ extra dimensions.
For example, for one extra dimension the right-handed neutrino
wavefunction spreads out over the extra dimension $R$, leading to
a suppressed Higgs interaction with the left-handed neutrino,
with a suppression factor of $1/\sqrt{M_{string}R}$.
This corresponds to the coupling between left and right-handed neutrinos
being more suppressed
for larger $R$, as the right-handed neutrino spends less
of its time on the 3 space dimensional brane where the left-handed
neutrino lives the larger $R$ becomes. The Dirac neutrino mass
is therefore suppressed relative to the electron mass, and may
be estimated as:
\be
m_{LR}^{\nu}\sim \frac{m_e}{\sqrt{M_{string}R}}
\sim \frac{M_{string}}{M_{Planck}} m_e
\ee
where we have used Eq.\ref{flat}. Clearly low string scales,
below the Planck scale, can lead to suppressed Dirac neutrino masses.
Similar suppressions can be achieved with anisotropic
compactifications \cite{Antusch:2005kf}.

\subsection{Cosmology}

What neutrinos lack in mass they make up for in number.
The Universe is filled with neutrinos, created less than a
second after the Big Bang, with each cubic centimetre of space
containing about 112 neutrinos of each species, giving more than
300 in total. This makes neutinos the second most abundant
particles in the Universe after the Cosmic Microwave Background (CMB) photons, where
each cubic centimetre of space contains 411 photons.
The CMB photons are the remnant of the Big Bang fireball,
originating from a time some 380,000 years after the Big Bang
when the Universe had cooled sufficiently to enable the first
atoms to form, thus rendering the Universe transparent.
By contrast there are on average a billion
times fewer electrons or protons that survived the great annihilation
of matter and antimatter which occurred about a second after the Big
Bang. Indeed
Cosmology today presents three major puzzles: why was there any
excess of matter over antimatter in the Universe; what is the major
matter constituent of the Universe; why is the Cosmological Constant
extremely small? Massive neutrinos may hold
important clues.

Matter and antimatter would have been created in equal amounts in the
Big Bang but all we see is a small amount of excess matter,
corresponding to about one electron or proton for every
billion photons (or neutrinos) in the Universe.
In order to generate matter-antimatter asymmetry in the Big Bang,
Sakharov in 1967
proposed a set of three necessary conditions: baryon number violation;
C and CP violation; and a violation of thermal equilibrium.
Perhaps surprisingly,
the Standard Model can satisfy of all these conditions,
for example baryon number is violated by non-perturbative
effects called sphalerons, but detailed calculations show that
it cannot lead to the desired matter-antimatter asymmetry
\cite{Shaposhnikov:1998vs}.
The see-saw
mechanism allows for a novel resolution to this puzzle. The idea, due
to Masataka Fukugita and Tsutomu Yanagida of Tokyo University
\cite{Fukugita:1986hr}, is that
when the Universe was very hot, just after the Big Bang, the heavy
right-handed neutrinos would have been produced, and could have decayed
preferentially into leptons rather than antileptons,
a possibility that is allowed since right-handed neutrinos
have Majorana masses that violate lepton number $L$,
neutrino interactions also may violate $CP$.
The excess leptons may subsequently be converted into an excess of
baryons via the Standard Model sphaleron effects mentioned above.
The see-saw mechanism therefore opens up the possibility of generating the
baryon asymmetry of the universe via ``leptogenesis''.
This process clearly requires CP violation for neutrinos, and
increases the motivation to discover leptonic CP violation
experimentally. However, as discussed, this would require Superbeams
or a Neutrino Factory.

\begin{figure}[t]
\centering
\includegraphics[width=0.76\textwidth]{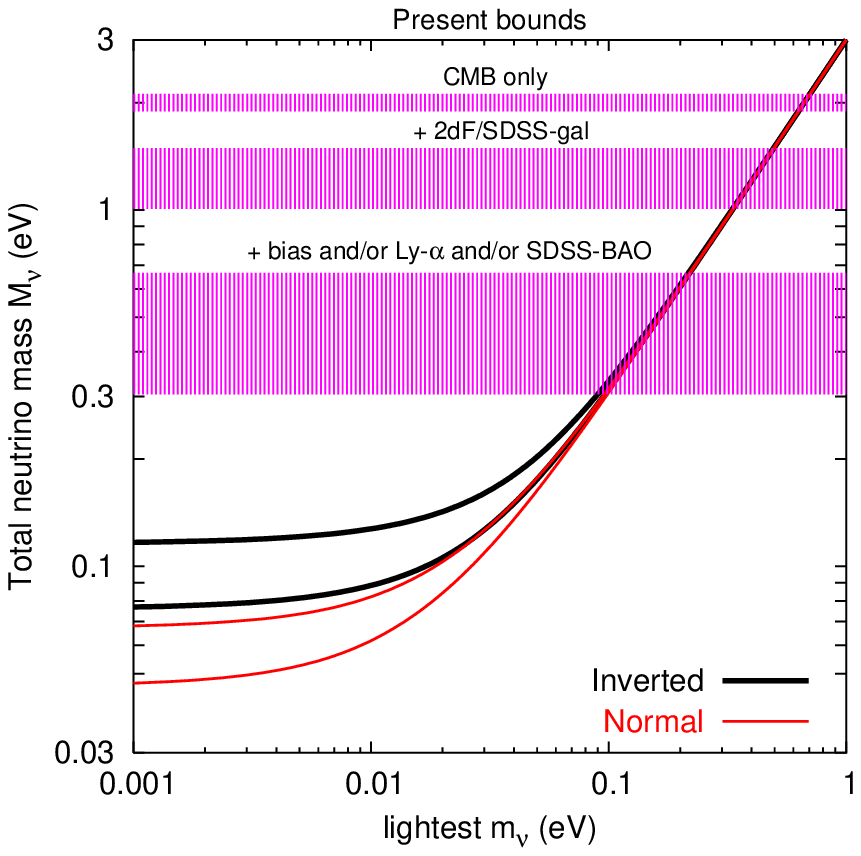}
\vspace*{-4mm}
    \caption{\footnotesize
The current bound on the sum of the three neutrino masses
at the 2 $\sigma $ level from cosmology using CMB only (upper band),
CMB combined with galaxy survey data (centre band), and then including other
effects (lower band) as discussed in \cite{Lesgourgues:2006nd} from where this
figure is taken. Also shown are the expected ranges for the sum
of the three neutrino masses as a function of the lightest neutrino
mass from neutrino oscillation data for both a normal mass
ordering (red lines) and an inverted mass ordering (black lines).
} \label{les}
\vspace*{-2mm}
\end{figure}

Studies of the kinematics of galaxies and galaxy clusters suggest that
at least 90 per cent of the mass of the Universe is made of
unknown dark matter.
Cosmology is sensitive to the absolute values of neutrino masses,
in the form of relic hot dark matter, where the dark matter is ``hot''
in the sense that the neutrinos were relativistic at the epoch
of galaxy formation. Such hot dark matter tends to lead to less
clumpiness of galaxy clusters, due to the free streaming
effects of such relativistic particles which tends to wash out
galaxy structures. On the other hand particles more massive than
neutrinos, known generically as weakly interacting massive particles
(WIMPs), can behave as cold dark matter (non-relativistic at the
epoch of galaxy formation) which tends to increase the
clumpiness of galaxy clusters. Such WIMPs can in principle be
directly detected in underground laboratories \cite{Morgan:2003ve}.

Recent results
from the Cosmic Microwave Background (CMB) experiments (especially
WMAP \cite{Peiris}) and galaxy redshift surveys (especially 2dF and SDSS \cite{Loveday:2002ax}),
gives a strong preference for cold dark matter over hot dark matter.
This leads to a limit on the amount of hot dark matter that can be accomodated.
When combined with oscillation
data, this leads to an upper limit on the absolute mass of each neutrino species of
about 0.3 eV \cite{Lesgourgues:2006nd}, corresponding to the
sum of neutrino masses being less than about 1 eV,
as shown in Fig.\ref{les}. More agresssive bounds are claimed
when other data are taken into account, also as shown in
Fig.\ref{les}. Neutrinos could constitute anything
from 0.1 to 2 per cent of the mass of the Universe,
corresponding to the heaviest
neutrino being in the mass range 0.05 to about 0.3 electronvolt.
Neutrinos any heavier than about 0.3 eV,
corresponding to the sum of neutrino masses exceeding about 1 eV,
would lead to galaxies being less clumped than actually
observed by the recent galaxy redshift surveys.
Ambitious claims are made that future CMB measurements
from the Planck satellite, due to be launched in 2008,
including the effect of weak gravitational lensing
(the deduction of large scale matter distributions from
CMB anisotropy) could constrain the sum of neutrino masses
down to 0.05 eV, corresponding to the atmospheric neutrino mass
in hierarchical models.
This illustrates the breathtaking rate at which neutrino
physics continues to advance.

\section{Conclusion}
\label{conclusions}

Since the discovery of neutrinos, just over half a century ago, we
have learned much about neutrinos, yet to a large extent neutrinos
remain somewhat elusive, if abundant, members of the Standard
Model families of quarks and leptons. Although we are not made of
neutrinos, the Universe as a whole is,
with over a billion neutrinos for every single atom.
Stars, such as the Sun, would not burn without neutrinos,
nor would supernovae explode, producing the star dust from which
we are made.

Almost a decade ago there was a revolution in neutrino physics
when they were found to have a tiny, but non-zero, mass,
in contradiction with the Standard Model.
It is fair to say that the past decade has been a golden age
of neutrino physics, with huge progress in neutrino physics both
on the experimental and theoretical fronts.
The surprise discovery of neutrino mass in atmospheric neutrinos,
was also accompanied by the further surprise of large, possibly
maximal, neutrino mixing. The solar neutrino ``problem'' is no more,
instead we have the discovery of solar neutrino ``mass'', again
involving large mixing.

These large mixings, large compared to
the corresponding quark mixing angles, can be understood from the
see-saw mechanism, for example by the sequential dominance mechanism,
but the see-saw mechanism is very difficult to test.
Nevertheless, if the see-saw mechanism in its simplest form
is correct, then it could explain the matter-antimatter
asymmety in the Universe via the leptogenesis mechanism.
However this would require both Majorana masses and CP
violation, neither of which have been experimentally established.
Alternatively, if neutrinos have Dirac masses, then their smallness
might be accounted for by invoking large extra dimensions.

Although neutrino physics has now entered the age of precision
measurements, much is left to learn. For example the CP violating
phase $\delta$ is yet to be measured, and the reactor angle
$\theta_{13}$ is similarly undetermined.
The neutrino mass ordering is not yet specified either,
nor is the absolute scale of neutrino mass,
and even the nature of neutrino mass itself has not been verified.
The answer to all these questions will
have profound implications for particle physics and
cosmology.

Given the recent results from MiniBooNE, which failed to confirm the LSND
signal, one might be tempted to think that the golden age of
major surprise discoveries in neutrino physics is over.
However it is just possible
that neutrino physics has further surprises up her sleeve.
For example, neutrinos may yet be observed to have
rather large Majorana masses, saturating
or even violating the cosmological bounds.
The recent controversial claim of a signal in neutrinoless
double beta decay is being vigorously checked,
and if confirmed would imply that neutrino masses have
quite a high degree of degeneracy.
This would certainly set the cat amongst the
cosmological and theoretical pigeons, and herald a new
neutrino revolution to rival the one described here.



\begin{thebibliography}{99}


\bibitem{Mohapatra:2006gs}
  R.~N.~Mohapatra and A.~Y.~Smirnov,
  Ann.\ Rev.\ Nucl.\ Part.\ Sci.\  {\bf 56} (2006) 569
  [arXiv:hep-ph/0603118].

\bibitem{Mohapatra:2005wg}
  R.~N.~Mohapatra {\it et al.},
  arXiv:hep-ph/0510213.


\bibitem{King:2003jb}
  S.~F.~King,
  Rept.\ Prog.\ Phys.\  {\bf 67} (2004) 107
  [arXiv:hep-ph/0310204].

\bibitem{Altarelli:2004za}
  G.~Altarelli and F.~Feruglio,
  New J.\ Phys.\  {\bf 6} (2004) 106
  [arXiv:hep-ph/0405048].

\bibitem{Mohapatra:2003qw}
R.~N.~Mohapatra,
``Understanding neutrino masses and mixings within the seesaw framework,''
arXiv:hep-ph/0306016.

\bibitem{bilenky}
For an up to date review of the history and current evidence for
neutrino oscillations see
W.M.Alberico and S.M.Bilenky, hep-ph/0306239, and references therein.

\bibitem{Murayama:2002xw}
  H.~Murayama,
  Phys.\ World {\bf 15} (2002) 35.

\bibitem{Bahcall:2004cc}
  J.~N.~Bahcall,
  arXiv:physics/0406040.

\bibitem{Pontecorvo:1957qd}
B.~Pontecorvo,
Sov.\ Phys.\ JETP {\bf 7} (1958) 172
[Zh.\ Eksp.\ Teor.\ Fiz.\  {\bf 34} (1957) 247];
B.~Pontecorvo,
Sov.\ Phys.\ JETP {\bf 26} (1968) 984
[Zh.\ Eksp.\ Teor.\ Fiz.\  {\bf 53} (1967) 1717].


\bibitem{MSW}
L. Wolfenstein, Phys. Rev. {\bf D17} (1978) 2369;
S. Mikheyev and A. Yu. Smirnov, Sov. J. Nucl. Phys. {\bf 42} (1985) 913.

\bibitem{MNS}
Z. Maki, M. Nakagawa and S. Sakata, Prog. Theo. Phys. {\bf 28} (1962)
247;
B.~W.~Lee, S.~Pakvasa, R.~E.~Shrock and H.~Sugawara,
Phys.\ Rev.\ Lett.\  {\bf 38} (1977) 937
[Erratum-ibid.\  {\bf 38} (1977) 1230].

\bibitem{SK} Y. Fukuda {\it et al.}  [Super-Kamiokande Collaboration],
Phys. Rev. Lett. {\bf 81}, 1562 (1998).

\bibitem{Ahmed:2003kj}
S.~N.~Ahmed {\it et al.}  [SNO Collaboration],
arXiv:nucl-ex/0309004.


\bibitem{Eguchi:2002dm}
K.~Eguchi {\it et al.}  [KamLAND Collaboration],
Phys.\ Rev.\ Lett.\  {\bf 90} (2003) 021802
[arXiv:hep-ex/0212021].

\bibitem{Maricic:2005sg}
  J.~Maricic and J.~G.~Learned,
  Contemp.\ Phys.\  {\bf 46} (2005) 1.


\bibitem{Ahn:2006zz}
  M.~H.~Ahn {\it et al.}  [K2K Collaboration],
  Phys.\ Rev.\  D {\bf 74} (2006) 072003
  [arXiv:hep-ex/0606032].

\bibitem{Michael:2006rx}
  D.~G.~Michael {\it et al.}  [MINOS Collaboration],
  Phys.\ Rev.\ Lett.\  {\bf 97} (2006) 191801
  [arXiv:hep-ex/0607088].

\bibitem{GoPe}
P.~C.~de Holanda and A.~Y.~Smirnov,
arXiv:hep-ph/0309299;
G.~L.~Fogli, E.~Lisi, A.~Marrone, D.~Montanino, A.~Palazzo and A.~M.~Rotunno,
arXiv:hep-ph/0310012;
A.~Bandyopadhyay, S.~Choubey, S.~Goswami and S.~T.~Petcov,
arXiv:hep-ph/0309236;
M.~Maltoni, T.~Schwetz, M.~A.~Tortola and J.~W.~Valle,
arXiv:hep-ph/0309130.

\bibitem{Maltoni:2004ei}
  M.~Maltoni, T.~Schwetz, M.~A.~Tortola and J.~W.~F.~Valle,
  New J.\ Phys.\  {\bf 6} (2004) 122
  [arXiv:hep-ph/0405172].

\bibitem{Apollonio:1999ae}
M.~Apollonio {\it et al.}
[CHOOZ Collaboration],
Phys.\ Lett.\ B {\bf 466} (1999) 415
[arXiv:hep-ex/9907037].


\bibitem{tribi}
P.~F.~Harrison, D.~H.~Perkins and W.~G.~Scott,
Phys.\ Lett.\ B {\bf 530} (2002) 167
[arXiv:hep-ph/0202074];
P.~F.~Harrison and W.~G.~Scott,
Phys.\ Lett.\ B {\bf 535} (2002) 163
[arXiv:hep-ph/0203209];
P.~F.~Harrison and W.~G.~Scott,
Phys.\ Lett.\ B {\bf 557} (2003) 76
[arXiv:hep-ph/0302025];
see also 
L.~Wolfenstein,
Phys.\ Rev.\ D {\bf 18} (1978) 958.


\bibitem{Athanassopoulos:1997pv}
C.~Athanassopoulos {\it et al.}  [LSND Collaboration],
Phys.\ Rev.\ Lett.\  {\bf 81} (1998) 1774
[arXiv:nucl-ex/9709006].


\bibitem{Eitel:2000by}
K.~Eitel  [KARMEN Collaboration],
Nucl.\ Phys.\ Proc.\ Suppl.\  {\bf 91} (2000) 191
[arXiv:hep-ex/0008002].

\bibitem{AguilarArevalo:2007it}
  A.~A.~Aguilar-Arevalo {\it et al.}  [The MiniBooNE Collaboration],
  Phys.\ Rev.\ Lett.\  {\bf 98} (2007) 231801
  [arXiv:0704.1500 [hep-ex]].


\bibitem{Itow:2001ee}
  Y.~Itow {\it et al.}  [The T2K Collaboration],
  arXiv:hep-ex/0106019.

\bibitem{Ayres:2004js}
  D.~S.~Ayres {\it et al.}  [NOvA Collaboration],
  arXiv:hep-ex/0503053.


\bibitem{Geer:1997iz}
  S.~Geer,
  Phys.\ Rev.\  D {\bf 57} (1998) 6989
  [Erratum-ibid.\  D {\bf 59} (1999) 039903]
  [arXiv:hep-ph/9712290].



\bibitem{Kraus:2004zw}
  C.~Kraus {\it et al.},
  Eur.\ Phys.\ J.\  C {\bf 40} (2005) 447
  [arXiv:hep-ex/0412056].

\bibitem{Osipowicz:2001sq}
A.~Osipowicz {\it et al.}  [KATRIN Collaboration],
arXiv:hep-ex/0109033.

\bibitem{Zuber:2004bx}
  K.~Zuber,
  Contemp.\ Phys.\  {\bf 45} (2004) 491.


\bibitem{Aalseth:2004hb}
  C.~Aalseth {\it et al.},
  arXiv:hep-ph/0412300.



\bibitem{Klapdor-Kleingrothaus:2001ke}
H.~V.~Klapdor-Kleingrothaus, A.~Dietz, H.~L.~Harney and I.~V.~Krivosheina,
Mod.\ Phys.\ Lett.\ A {\bf 16} (2001) 2409
[arXiv:hep-ph/0201231].

\bibitem{Feruglio:2002af}
F.~Feruglio, A.~Strumia and F.~Vissani,
arXiv:hep-ph/0201291.



\bibitem{Aalseth:2002dt}
C.~E.~Aalseth {\it et al.},
arXiv:hep-ex/0202018.

\bibitem{Klapdor-Kleingrothaus:2002kf}
H.~V.~Klapdor-Kleingrothaus,

\bibitem{KlapdorKleingrothaus:2004wj}
  H.~V.~Klapdor-Kleingrothaus, I.~V.~Krivosheina, A.~Dietz and O.~Chkvorets,
  Phys.\ Lett.\  B {\bf 586} (2004) 198
  [arXiv:hep-ph/0404088].

\bibitem{SNO++}
This proposal is so new that the only reference I could find
is in Wikipedia. Watch this space.


\bibitem{seesaw}
P.~Minkowski,
  Phys.\ Lett.\ B {\bf 67} (1977) 421;
M. Gell-Mann, P. Ramond and R. Slansky in Sanibel Talk,
CALT-68-709, Feb 1979, and in {\it Supergravity} (North Holland,
Amsterdam 1979);
T. Yanagida in {\it Proc. of the Workshop on Unified Theory and
Baryon Number of the Universe}, KEK, Japan, 1979;
S.L.Glashow, Cargese Lectures (1979);
R.~N.~Mohapatra and G.~Senjanovic,
Phys.\ Rev.\ Lett.\  {\bf 44} (1980) 912;
J.~Schechter and J.~W.~Valle,
Phys.\ Rev.\ D {\bf 25} (1982) 774.


\bibitem{Antusch:2004gf}
  S.~Antusch and S.~F.~King,
  New J.\ Phys.\  {\bf 6} (2004) 110
  [arXiv:hep-ph/0405272].

\bibitem{King:2006hn}
  S.~F.~King,
  arXiv:hep-ph/0610239.

\bibitem{King:1998jw}
  S.~F.~King,
  Phys.\ Lett.\  B {\bf 439} (1998) 350
  [arXiv:hep-ph/9806440].

\bibitem{King:1999mb}
  S.~F.~King,
  Nucl.\ Phys.\  B {\bf 576} (2000) 85
  [arXiv:hep-ph/9912492].
  S.~F.~King,
  JHEP {\bf 0209} (2002) 011
  [arXiv:hep-ph/0204360].

\bibitem{King:2005bj}
  S.~F.~King,
  JHEP {\bf 0508} (2005) 105
  [arXiv:hep-ph/0506297].


\bibitem{Antoniadis:2005aq}
  I.~Antoniadis,
  arXiv:hep-ph/0512182.


\bibitem{Arkani-Hamed:1998vp}
N.~Arkani-Hamed, S.~Dimopoulos, G.~R.~Dvali and J.~March-Russell,
Phys.\ Rev.\ D {\bf 65} (2002) 024032
[arXiv:hep-ph/9811448].

\bibitem{Antusch:2005kf}
  S.~Antusch, O.~J.~Eyton-Williams and S.~F.~King,
  JHEP {\bf 0508} (2005) 103
  [arXiv:hep-ph/0505140].

\bibitem{Shaposhnikov:1998vs}
  M.~E.~Shaposhnikov,
  Contemp.\ Phys.\  {\bf 39} (1998) 177.



\bibitem{Fukugita:1986hr}
  M.~Fukugita and T.~Yanagida,
  Phys.\ Lett.\  B {\bf 174} (1986) 45.



\bibitem{Morgan:2003ve}
  B.~Morgan,
  Contemp.\ Phys.\  {\bf 44} (2003) 219.

\bibitem{Peiris}
  H.~Peiris,
  Contemp.\ Phys.\  {\bf 46} (2005) 77.


\bibitem{Loveday:2002ax}
  J.~Loveday  [the SDSS Collaboration],
Contemp.\ Phys.\  {\bf 43} (2002) 437.
  arXiv:astro-ph/0207189.



\bibitem{Lesgourgues:2006nd}
  J.~Lesgourgues and S.~Pastor,
  Phys.\ Rept.\  {\bf 429} (2006) 307
  [arXiv:astro-ph/0603494].





\end{thebibliography}
\end{document}